\documentclass[aps,pra,reprint,showpacs]{revtex4-1}

\usepackage{graphicx}
\usepackage{amsfonts}
\usepackage{amssymb}
\usepackage{amsmath}

\newcommand{\be}{\begin{equation}}
\newcommand{\ee}{\end{equation}}
\newcommand{\bea}{\begin{eqnarray}}
\newcommand{\eea}{\end{eqnarray}}

\begin{document}

\title{Dysprosium magneto-optical traps}

\author{Seo Ho Youn}
\author{Mingwu Lu}
\author{Ushnish Ray}
\author{Benjamin L. Lev}
\affiliation{Department of Physics, University of Illinois at Urbana-Champaign, Urbana, IL 61801-3080 USA}

\begin{abstract}
Magneto-optical traps (MOTs) of highly magnetic lanthanides open the door to explorations of novel phases of strongly correlated matter such as lattice supersolids and quantum liquid crystals.  We recently reported the first MOTs of the five high abundance isotopes of the most magnetic atom, dysprosium.  Described here are details of the experimental technique employed for repumper-free Dy MOTs containing up to half a billion atoms.  Extensive characterization of the MOTs' properties---population, temperature, loading, metastable decay dynamics, trap dynamics---is provided.  
\end{abstract}
\date{\today}
\maketitle

\section{Introduction}

Ultracold gases of fermionic dipolar atoms provide the opportunity to examine---in a pristine and tunable setting---the non-Fermi liquid, strongly correlated electronic behavior manifest in some of the most interesting materials of late:  high-\emph{T}$_c$ cuprate superconductors, strontium ruthenates, 2D electron gases, and iron-based superconductors.  Strong and competing interactions in these systems induce transitions to states beyond the familiar insulating, metallic, and superconducting.  Specifically, phases that break rotational and translational symmetries emerge in a manner akin to those found in classical liquid crystals, e.g., the nematic and smectic~\cite{Fradkin2009,Fradkin2010}.  Although quantum liquid crystal (QLC) theory can describe these non-Fermi liquids in a compelling and general framework, unwanted solid state material complexity---disorder and dynamical lattice distortions---can obscure the underlying electronic physics, and lack of wide system tunability can hamper efforts to fully explore QLC phase portraits.

Exploiting large dipole-dipole interactions (DDI) in ultracold gases will allow the exploration of QLC physics in this inherently more tunable and characterizable system.  Recent theoretical proposals employing strongly magnetic fermionic atoms or polar fermionic molecules have begun to shed light on the accessible QLC physics.  These include predictions of uniaxial (meta-nematic)~\cite{Miyakawa:2008} and biaxial nematic~\cite{Fregoso:2009} distortions of the Fermi surface in the presence of a polarizing field, and meta-nematic and smectic phases in 2D anisotropic optical lattices~\cite{Quintanilla:2009}.  An exciting prospect lies in the possibility of observing spontaneous magnetization in dipolar systems, and Refs.~\cite{Fregoso2009b,fregoso:2010} postulate the existence of observable quantum ferro-nematic phases and spin textures in ultracold highly magnetic fermionic atomic gases in zero polarizing field.

While many exciting results will continue to arise from the degenerate fermionic dipolar molecule system~\cite{Doyle04EPJDreview,Ye:2009,Ospelkaus2010,Ni2010}, to most easily observe true (non-meta) QLC phases, the symmetries of interest should be spontaneously broken, which is not possible when employing ground state polar molecules.  This is because the strong, $r^{-3}$ character of the DDI is realized only in the presence of a rotational symmetry breaking, polarizing electric field that mixes opposite parity states.  Ultracold chemical reactions can also hamper the use of fermionic polar molecules for studies of DDI-induced exotic phases in 3D, as recent experiments in KRb have shown~\cite{Ospelkaus2010,Ni2010}.  In contrast, highly magnetic atoms exhibit the DDI interaction even in the absence of a polarizing field and are largely immune to chemical and inelastic collisions when spin-polarized in an optical dipole trap.  While both the electric and magnetic DDI can be continuously tuned to zero, only the magnetic DDI can be tuned negative~\cite{Pfau02}.  

One must look to the rare earth (lanthanide) series to find atoms possessing masses and magnetic moments large enough to support exotic bosonic and fermionic phases.  The extraordinarily large magnetic dipole of dysprosium (10 $\mu_{B}$), which possesses the largest magnetic moment of any fermionic atom---and is tied with terbium in possessing the largest moment among bosonic atoms~\footnote{Terbium has only one isotope, a boson.  Unfortunately, it possesses a 400 K electronic state that could be driven by incoherent blackbody radiation, thus limiting coherence and trap lifetimes.  Additionally, thulium (4 $\mu_{B}$)~\cite{Thulium2010} and holmium (9 $\mu_{B}$) have only single bosonic isotopes.}---is likely sufficient to induce QLC phases~\cite{Quintanilla:2009,Fregoso2009b,fregoso:2010}.  We describe here in detail the Dy magneto-optical traps (MOTs)~\cite{Lu2010} that may serve as a precursor to degenerate, highly magnetic Fermi---and Bose---gases. 

It is not yet clear whether a degenerate gas of highly magnetic lanthanide atoms is possible to create, but overcoming a large DDI while cooling to degeneracy has precedent in the Bose-Einstein condensation (BEC) of chromium~\cite{Pfau05CrBEC}.  The attainment of Cr BEC opened the door to bosonic ultracold dipolar physics~\cite{PfauReview09}.   Chromium possesses the large magnetic dipole moment of 6 Bohr magnetons ($\mu_B$).  This is 6-times larger than that of the alkali atoms'---Rb has a magnetic moment of $\mu=1$ $\mu_B$ in the doubly-polarized state---and represents a significant 36-fold enhancement in the DDI strength \be U_{dd}=\frac{\mu_{0}\mu^{2}}{4\pi}\frac{1-3\cos^{2}\theta}{|\mathbf{r}|^3},\ee where $r$ is the distance between two dipoles and $\theta$ is the angle between the direction of polarization and the relative position between the two particles.  However, as strong as Cr's DDI may be, recent calculations~\cite{Yi:2007} suggest that novel lattice phases predicted by the extended Bose-Hubbard (eBH) model~\cite{PfauReview09}, such as density waves and lattice supersolids~\cite{Duan2010}, lie just beyond the reach of Cr's capability.   Specifically, the DDI energy must dominate the contact interaction energy, which occurs when $\epsilon=\mu_{0}\mu^{2}m/12\pi\hbar^2a_{s}\gtrsim1$, where $m$ is the mass, $a_{s}$ is the s-wave scattering length, and the extra factor of 3 in the denominator designates $\epsilon\geq1$ as the regime in which homogeneous dipolar BECs become unstable to collapse~\cite{PfauReview09}.  To observe novel lattice phases, $\epsilon$ should be $>$0.7--0.8~\cite{Yi:2007}.  For $^{52}$Cr, $\epsilon_{\text{Cr}}=0.15$, and even with the demonstrated five-fold reduction of $a_{s}$ via a Feshbach resonance~\cite{Pfau07CrBECferrofluid}\footnote{Rapid three-body losses preclude greater reduction of $a_{s}$~\cite{Pfau07CrBECferrofluid}}, $\epsilon_{\text{Cr}}=0.7$ remains at or below the threshold for new phases.  In contrast, $\epsilon_{\text{Dy}}=1.34$ which is 9$\times$ larger than $\epsilon_{\text{Cr}}$, assuming that the as yet unmeasured scattering length for at least one of Dy's bosonic isotopes is approximately equal to $^{52}$Cr's $a_{s}=100$ $a_{0}$.  With such a large $\epsilon_\text{Dy}$, exploring supersolids and density wave phases without the use of Feshbach resonances should be possible with Dy.

As candidates for fermionic dipolar physics, existing MOTs of highly magnetic fermionic $^{53}$Cr and $^{167}$Er are not yet populous enough to contemplate cooling to degeneracy~\cite{Mcclelland:2006,FrenchCr06}.  The technique of buffer gas cooling has been used for cooling lanthanides and has proven successful in producing large 500 mK samples~\cite{Hancox:2004}.  While Dy has been adiabatically cooled to $\sim$50 mK at final densities of $\sim$$10^{9}$ cm$^{-3}$~\cite{Newman2010} in such an experiment, evaporative cooling in a magnetic trap to lower temperatures has not yet proven effective due to large inelastic collisions.  Thus, the extraordinarily populous 0.1--2 mK Dy samples discussed here open the door to a rich landscape of physics.

\begin{figure}[t]
\includegraphics[width=0.49\textwidth]{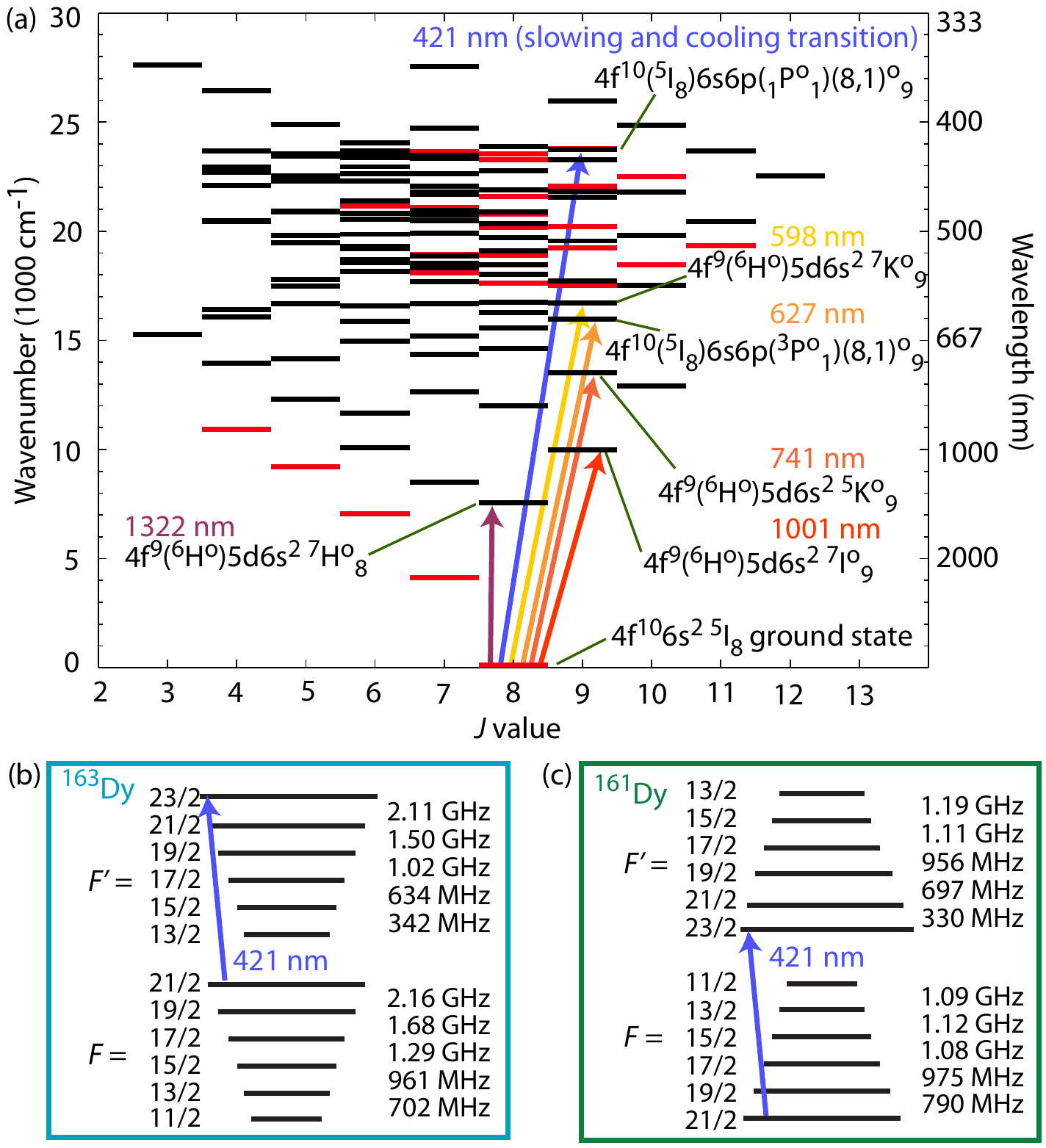}
\caption{(color online).  Dy energy level structure~\protect\cite{Martin:1978,Leefer:2009b,Hancox:2005}, where $J$ is the total electronic angular momentum and $F=J+I$, the total electronic plus nuclear angular momentum.  The prime indicates excited state quantum numbers.  (a) The MOT and Zeeman slower use the strongest laser cooling transition at 421 nm; there are nearly 140 lower energy metastable states.  Several other laser cooling transitions exist between the ground state and the excited states.  The red states are even parity and the odd are shown in black.  Dysprosium has five high abundance isotopes; three bosons ($^{164}$Dy, $^{162}$Dy, and $^{160}$Dy with $I=0$) and two fermions ($^{163}$Dy and $^{161}$Dy with $I=5/2$).  (b--c) The two fermions have oppositely signed nuclear spin, resulting in a relative inversion in the hyperfine structure.}
\label{fig:dy_levels}
\end{figure}

Ultracold samples of bosonic and fermionic isotopes of Dy would find application beyond the QLC and eBH physics described above.   Co-trapping isotopes of Dy present an interesting system to explore Bose-Fermi mixtures of near equal mass, reminiscent of $^3$He-$^4$He studies, but now in the presence of same and cross-species dipolar interactions.  Studies of large-spin degenerate spinor gases~\cite{Stamper-Kurn:2006,Wu:2006}, simulations of dense nuclear matter~\cite{Baym09}, creation of unconventional superfluid pairing~\cite{CWu2010}, and explorations of zero sound~\cite{Bohn10} and roton modes~\cite{Bohn07} in dipolar gases are exciting avenues of research.  In addition, ultracold samples of Dy will aid precision measurements of parity nonconservation and variation of fundamental constants~\cite{Leefer:2009}, single-ion implantation~\cite{Mcclelland:2006}, and quantum information processing~\cite{Derevianko:2004,Saffman:2008}.  With regard to the latter, the low-lying telecom (1322 nm) and InAs quantum dot amenable (1001 nm) transitions (see Fig.~\ref{fig:dy_levels}) will be useful for creating hybrid atom-photonic or atom-quantum systems that exploit potentially long-lived spin coherences in ensembles of Dy for quantum memory.  Novel ultracold collisions~\cite{Connolly2010} and complex Feshbach resonance-induced molecular association phenomena are expected to appear in traps of these non-$S$ state atoms, which may enable exploration of 1D strongly correlated gases with highly magnetic molecules~\cite{Zoller2010}.

\section{Laser cooling system}\label{lasercooling}

\begin{figure}[h]
\includegraphics[width=0.49\textwidth]{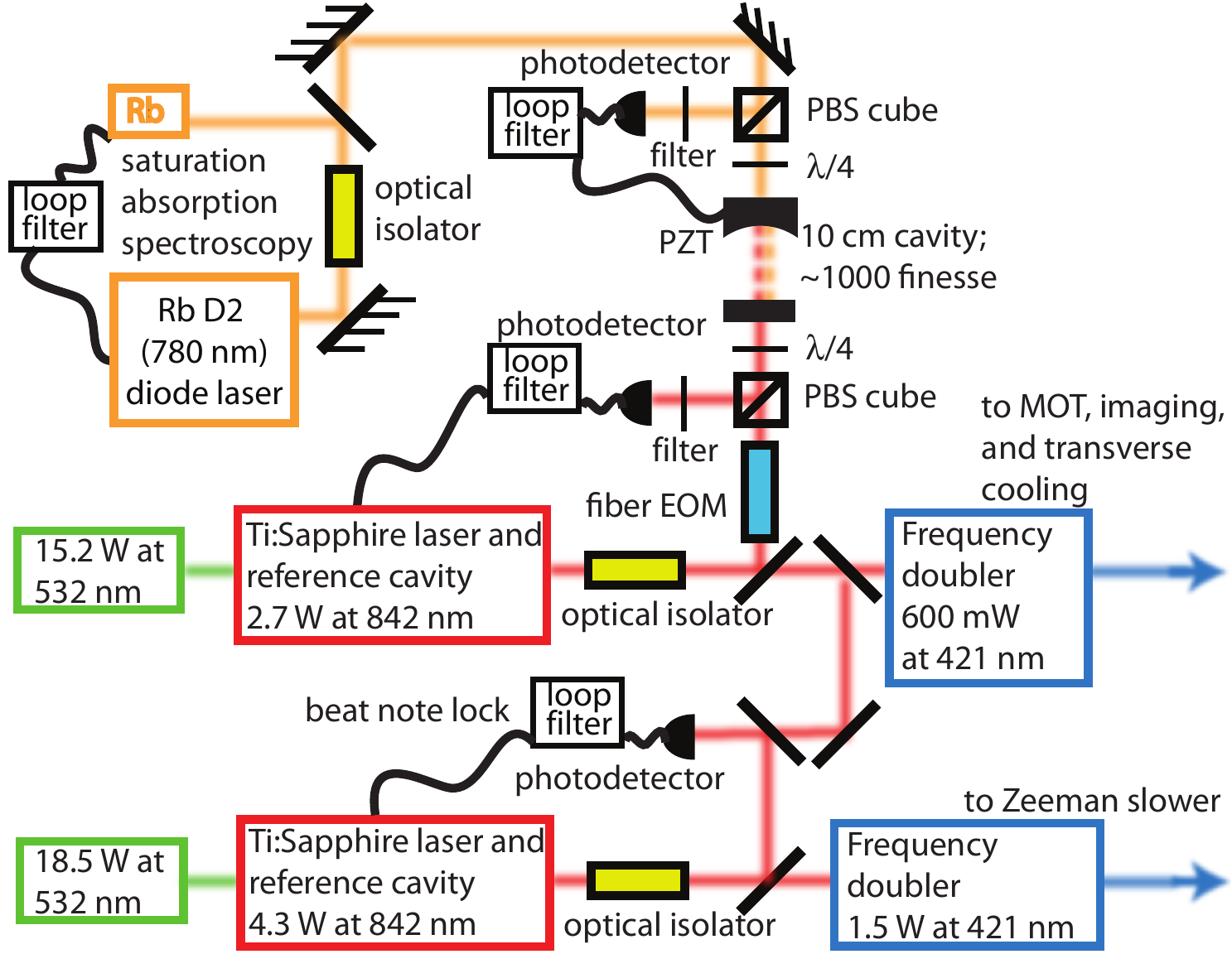}
\caption{\label{LaserSetUp}(color online).  Schematic of the Dy laser cooling system.  See text for details.}
\end{figure}

A simple laser cooling and trapping method for highly magnetic atoms has now proven successful for three magnetic lanthanides, Er, Dy, and Tm~\cite{Mcclelland:2006,Lu2010,Thulium2010}.  Measurements of MOT recycling dynamics show that the working principle behind the Er and Dy MOTs are similar~\cite{Mcclelland:2006,Lu2010}.  Despite the existence of more than a hundred levels between ground and the open excited state of the cooling transition, no repumping laser is necessary because of a novel recycling mechanism~\cite{Mcclelland:2006}:  After decaying out of the MOT cooling transition to metastable states, the highly magnetic atoms remain confined in the MOT's magnetic quadrupole trap while the metasable state population recycles to the ground state, at which point the atom are recaptured by the MOT.  The strongest transition $J\rightarrow J+1$ transition is used for Zeeman slowing and MOT cooling.  Whereas weakly magnetic atoms would be lost from the trapping region during time in which the atoms spend in the metastable states, the strong dipole moments (7 $\mu_{B}$ for Er) allow confinement even in the MOT's magnetic quadrupole gradient.  The 1:10$^5$ branching ratio between population decay back to the ground state versus to the metastable state is sufficient for a MOT to form given the efficient population recycling.  

We highlight four criteria necessary for the large-population, highly magnetic, repumperless MOT to work when using strong open transitions:  1) to be captured by the MOT, the branching ratio must be small enough that atoms exit the Zeeman slower in their ground state;  2) the decay channel through the metastable states must be rapid enough that inelastic collision processes---from background, spin-changing, or light induced collisions---do not deplete the metastable magnetic trap (MT) reservoir; 3) decay through the metastable MT should be slow enough for it to serve as a reservoir to capture more atoms than those cycling on the MOT transition; 4) for isotopes with $I\neq0$, the difference frequencies between adjacent $F\rightarrow F'=F+1$ hyperfine transitions in the ground and excited states must be small enough that the cooling laser itself serves as a repumper (see Sec.~\ref{shelving}).

All of these criteria are satisfied in the five most abundant isotopes of Dy, the level structures of which are depicted in Fig.~\ref{fig:dy_levels}.  Before discussing the Dy MOT recycling dynamics and population in Sec.~\ref{MOTcharacter}, we first describe the laser system employed for producing the light necessary to drive the Dy laser cooling transition at 421 nm.    

The strongest  $J\rightarrow J+1$ cycling transition is the 421-nm line, which has a broad linewidth of $31.9(0.7)$ MHz~\cite{Martin:1978,Lu2010b}.  
A total of 2 W of 421-nm optical power is generated by two Ti:sapphire laser systems.  Such high power is necessary because the saturation intensity 58~mW/cm$^{2}$ is large and, as we discuss below, the Zeeman slower optimally uses 1 W of power.  The first Ti:sapphire system (TiS1) is pumped with 15.2 W at 532 nm, which when frequency doubled in a ring cavity with an LBO crystal produces 600 mW of continuous wave (CW) 421-nm light with a 20-kHz linewidth~\cite{Tekhnoscan}.  The second doubled Ti:sapphire laser system (TiS2) produces up to 1.6~W CW at 421 nm (again using a ring cavity frequency doubler) from a 4.3 W 842-nm beam. TiS2 is pumped by a 18.5 W 532-nm laser, and the 421-nm light has a linewidth of $\sim$50 kHz at 421 nm.  

TiS1 produces the 421-nm light for the MOT, transverse cooling, and imaging beams, while TiS2 is used solely for the Zeeman slower.  Each laser is locked, at 842 nm, to its own low-finesse reference cavity.  These cavities exhibit $\sim$40 MHz/hr frequency drift, so we have developed a more stable locking scheme which provides a long-term stability of several 100 kHz.  The frequency reference is derived using the transfer cavity technique:  the 842-nm light from TiS1 is locked to a mode of an optical cavity whose length is stabilized by a 780-nm laser that is resonant with the cavity $\sim$$2\times10^{4}$ modes away.  This 780-nm laser is itself frequency stabilized to a hyperfine transition in Rb using saturation-absorption spectroscopy.  Thus, the stability of a medium-finesse cavity locked to a Rb frequency reference is transferred to the 842-nm laser of the TiS1 system.  The TiS2 system is offset locked~\cite{schunemann_simple_1999} at the Zeeman slower laser detuning by detecting a beat note at 842-nm between the TiS1 and TiS2 lasers.  See Appendix~\ref{laserlock} for more details.
\begin{figure}[t]
\includegraphics[width=0.49\textwidth]{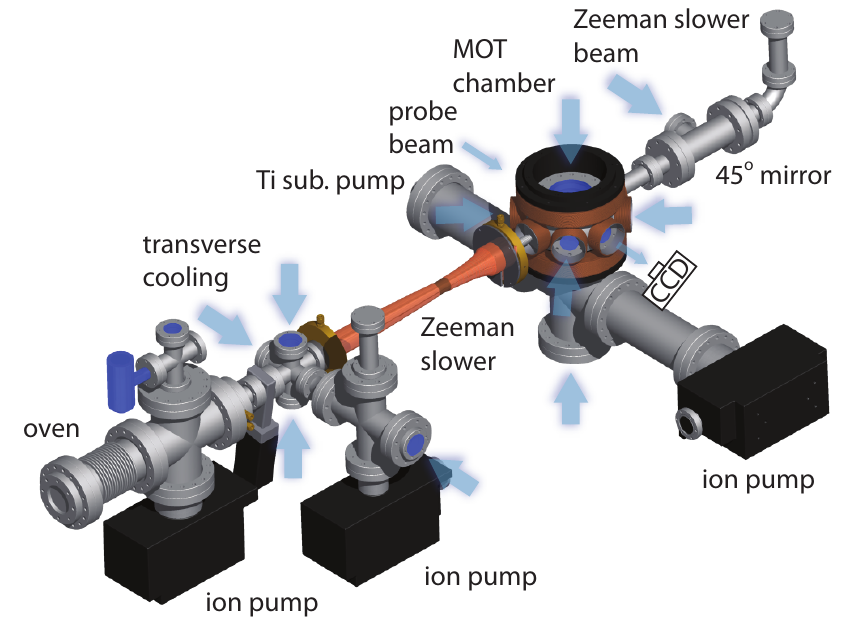}
\caption{\label{DyVacuumChamber}(color online).  Sketch of the Dy UHV chamber system. The blue arrows depict the slowing, cooling, and imaging laser beams.}
\end{figure}

\section{Trapping apparatus}

Fig.~\ref{DyVacuumChamber} illustrates the schematic of our Dy vacuum chamber system.  A thermal atomic beam is first generated with a high-temperature oven.  This atomic source is then collimated through a differential pumping tube, and a larger flux is achieved by transverse cooling.  The Dy atoms are decelerated and cooled by a ``spin-flip'' Zeeman slower.  Finally, the slowed atoms are captured and loaded into a large-gradient MOT in the trapping chamber.

The vacuum system consists of four main sections: high-temperature oven, transverse cooling, Zeeman slower, and MOT trapping chamber.  Each section---except the Zeeman slower---has an ion gauge to monitor the vacuum pressure and a dedicated ion pump.  In addition, there is a stainless steel tube (inner diameter 4.6 mm, length 18 cm) that provides differential pumping between the oven and transverse cooling sections.  The Zeeman slower (inner diameter 1.7 cm, length 54 cm) also serves as a differential pumping tube between the transverse cooling and MOT sections.  A titanium sublimator provides additional pumping in the MOT trapping chamber.

\subsection{High-temperature oven and transverse cooling}

Because Dy has a very high melting point of 1412 $^\circ$C, a dual-filament, all tantalum high-temperature effusion cell with a water-cooling shroud~\footnote{Custom made from SVT Associates, Inc.} is used to heat mm-sized pieces of Dy.  The non-isotopically purified  Dy is placed in the tantalum crucible and heated to 1250 $^\circ$C during typical MOT operation.  A near-uniform temperature is achieved by two servos controlling heaters located at the the tip and main part of the oven.  The crucible with a 5-mm diameter orifice generates a Dy beam.  By attaching a flexible bellow to the oven and controlling it with mechanical positioning stages, the direction of atomic beam is optimized for maximum MOT population.  

Because blackbody radiation from the crucible can heat the UHV chamber walls, a 2.25" diameter tantalum shield is installed around the tip of the differential pumping tube, immediately before the transverse cooling section.  Heat propagation through the vacuum system is avoided by continuously water cooling the ConFlat (CF) flange to which the tantalum disc and differential pumping tube are welded. The atomic beam may be shuttered by an in-vacuum, pneumatically controlled tantalum shield positioned a few mm from the crucible orifice.

Typical oven operation at 1250 $^\circ$C provides a Dy vapor pressure inside the crucible of $\sim$$7\times10^{-2}$ Torr. We achieve a vacuum of $1.0\times10^{-9}$ Torr at the ion gauge of the oven section with the use of a 75 L/s ion pump.  This low pressure is achieved with help from the getter properties of the evaporated Dy on the tantalum disk~\cite{li_reduction_2002}.  Fifteen grams of Dy lasts $\sim$400 h at 1250 $^\circ$C operation.  A gate valve after the oven section maintains the vacuum in the rest of system while detaching the oven for the refilling.  At the transverse cooling stage, the pressure is $8.2\times10^{-11}$ Torr when continuously evacuated with a 55 L/s ion pump.  

A larger atomic beam flux is achieved by transverse cooling the portion of the beam after the differential pumping tube and before the Zeeman slower.  A detailed study of Dy transverse cooling appears in Ref.~\cite{Leefer:2010}.  The transverse cooling beams are elliptically shaped such that they match the atomic beam; their dimensions are 4.4 mm by 18 mm.  One beam is oriented horizontally to the the atomic beam and is retroreflected through the chamber via anti-reflection coated windows.  The second beam enters the chamber vertically, passes through a quarter-wave plate, into the chamber and is again retroreflected through a quarter-wave plate before entering the chamber.  Addition of the quarter-wave plates along the vertical direction enhances transfer efficiency; no such gain is observed for the horizontal branch, presumably since the local magnetic field is nonzero.  The cooling beams are all red-detuned from the 421-nm transition by 0.2--0.4~$\Gamma$.  At a total power of $\sim$200 mW divided among the horizontal and vertical beams, the MOT population is enhanced by up to a factor of 4.  Additional power slightly reduces the MOT population, ostensibly due to increased metastable population shelving before the Zeeman slower (and possibly recoil heating).  See Sec.~\ref{shelving} for more details.

\begin{figure}[t]
\includegraphics[width=0.49\textwidth]{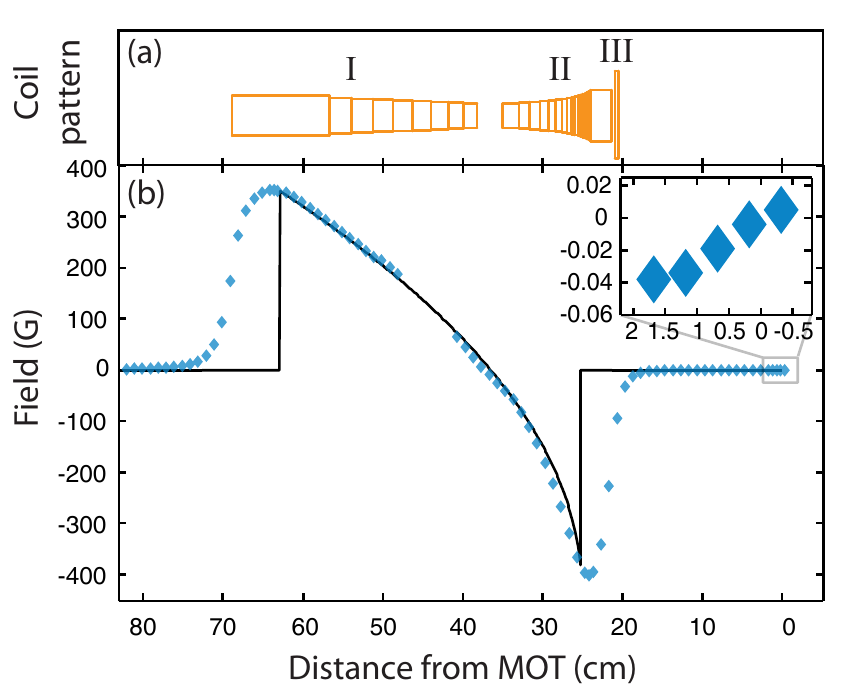}
\caption{\label{ZSlowerProfile}(color online).  Magnetic field profile of Zeeman slower as a function of position away from the center of the trapping chamber. Target (line) and measured ($\diamond$) fields are shown. The profile is measured with 3.01 A in Section I, 3.02 A in Section II, and 40.24 A in Section III. Note that Zeeman slower was designed for 50$\%$ maximum deceleration efficiency.  The optimal current for MOT production is 1.08 A in Section I, 2.66 A in Section II, and 40.21 A in Section III. Inset shows field zeroing at MOT position.}
\end{figure}

\subsection{Zeeman slower}

The spin-flip (zero-crossing) Zeeman slower~\cite{MetcalfBook99} decelerates and cools the collimated atomic beam from 480 m/s (most probable velocity) to 30 m/s~\footnote{These are design values, and have not been measured.}.  The inhomogeneous magnetic field profile of the Zeeman slower is shown in Fig.~\ref{ZSlowerProfile}, and the positive and the negative fields are generated from three distinct coils.  The first section, which is closest to the oven, has 11 staggered layers of magnet wire helically wound around the 1.1'' CF nipple forming the Zeeman slower.  The second section has 18 layers with current flowing in the opposite direction to the first.  A third section is added to cancel the stray field from the first two sections at the MOT position.  It consists of a 10-turn coil of hollow square magnet wire (4.2 mm on a side) through which chilled water circulates.  Because of the heat dissipated in the first two solenoid sections, chilled water through copper tubing wrapped around them is necessary to keep the temperature below 30 $^\circ$C. 

\begin{figure}[t]
\includegraphics[width=0.49\textwidth]{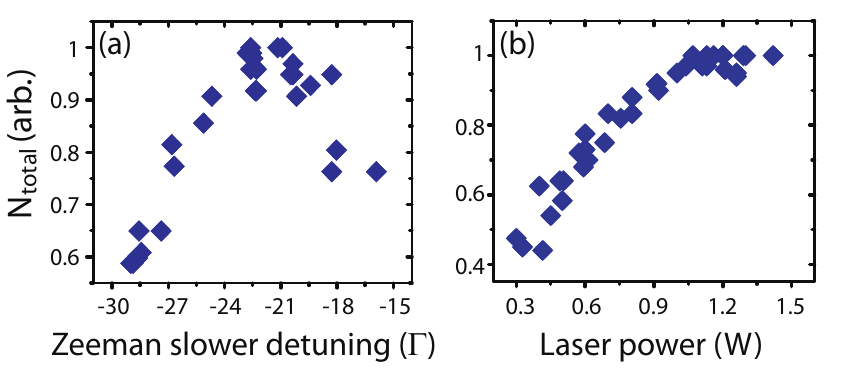}
\caption{\label{NVsZSbeam}(color online).  MOT population as a function of (a) Zeeman slower detuning and (b) Zeeman slower laser power. Optimal operation occurs for a detuning of -21 $\Gamma$.  Population saturates above 1 W of laser power.}
\end{figure}

Up to 1.5 W of Zeeman slower beam power (Fig.~\ref{NVsZSbeam}) is used to slow the atomic beam.  The laser is detuned -21 $\Gamma$ [Fig.~\ref{NVsZSbeam}(a)] from the 421-nm atomic transition~\footnote{Unless stated, all the values of detuning are with respect to $^{164}$Dy atomic transition.  Other isotopes require slightly modified detuning for optimal MOT performance.} and the beam is focused into the orifice of the oven crucible (see Fig.~\ref{DyVacuumChamber}).  A 45$^\circ$ aluminum mirror is installed inside the vacuum system to avoid Dy coating the entrance viewport of the Zeeman slower laser beam.  The Dy film coating the mirror is reflective at 421 nm.  The diameter of the slowing laser is 2.5 cm at the vacuum window entrance.  As shown in Fig.~\ref{NVsZSbeam}(b), total trapped atom population $N_{\text{Total}}$ saturates at $\sim$1 W of input power; presumably power broadening aids the velocity capture range of the slower.

\subsection{MOT trapping chamber}

The MOT employs a three-retroreflected-beam configuration (a six-beam configuration can also be used, with similar results).  Each beam is aligned and collimated in free-space with a waist~\footnote{All beam waists are reported as a beam $1/e^2$ radius.} of 1.1 cm, and the MOT typically has a detuning of $\Delta=-1.2$ $\Gamma$.  A total intensity of 0.17--0.2$I_\text{s}$ provides the maximum MOT population, where $I_\text{s} \approx 2.7\times58$ mW/cm$^2$~\footnote{The additional factor of 2.7 accounts for approximately isotropic polarization and equally distributed $m_{J}$'s in the MOT.}.  A stainless steel octagon chamber with two 6'' diameter CF viewports on top and bottom and six 2.75'' CF viewports on the side provide the optical access necessary for the MOT beams and imaging beam.  The remaining two CF ports provide access for the atomic beam and counterpropagating Zeeman slower beam.

The magnetic quadrupole field used for both the MOT and magnetic trap~\cite{Lu2010} is generated from a coil pair in near-anti-Helmholtz configuration.  The coils generate $\nabla_{z}\text{B}=0.69\,{\rm G/cm\cdot A}$ along $\hat{z}$, which points along the quadrupole axis of symmetry; each coil has a cross-section of $10$ rows ($\hat{z}$) and $7$ columns ($\hat{\rho}$).  The coils are water cooled to support the $\sim$30 A used for the MOT.  The electric current is controlled by a servo providing a 440 $\mu$s turn-off time.  Stray field cancellation coils reduce the residual field to $\leq1.8$ mG.  

We achieve a MOT chamber pressure of typically $1.2\times10^{-11}$ Torr during MOT operation.  The vacuum is created with the help of a 75 L/s ion pump and a titanium sublimation pump.  We have noticed a lowering of the vacuum pressure from the evaporated Dy after the oven has run for several hours.  

\subsection{Imaging}

After the atoms are released from the trap, absorption images are taken from the CCD camera installed on the side of the trapping chamber with a probe beam pulse width of $60$ $\mu$s.  This short pulse width ensures that that atoms are approximately stationary and shelving into metastable states does not occur during the imaging process.  In order to take reliable temperature measurements of the MOT while avoiding eddy currents (lasting $\sim$1.5 ms) incurred during magnetic quadrupole coil shut-off (at $t_{d}=0$), the time-of-flight data were taken in 500 $\mu$s steps from $t_{d}=3$ to 5--7 ms.

\section{Dy MOT characteristics}\label{MOTcharacter}

In this section we discuss a detailed, semiclassical model and supporting data for how the repumper-less Dy MOT functions.  Measurements of the Dy MOT population and temperature versus MOT loading time, intensity, detuning, and magnetic quadrupole field gradient are presented. We also compare Dy MOT characteristics to those of the Er MOT~\cite{Mcclelland:2006,Berglund:2007}.  

\subsection{MOT decay and motional dynamics}

\begin{figure}[t]
\includegraphics[width=0.3\textwidth]{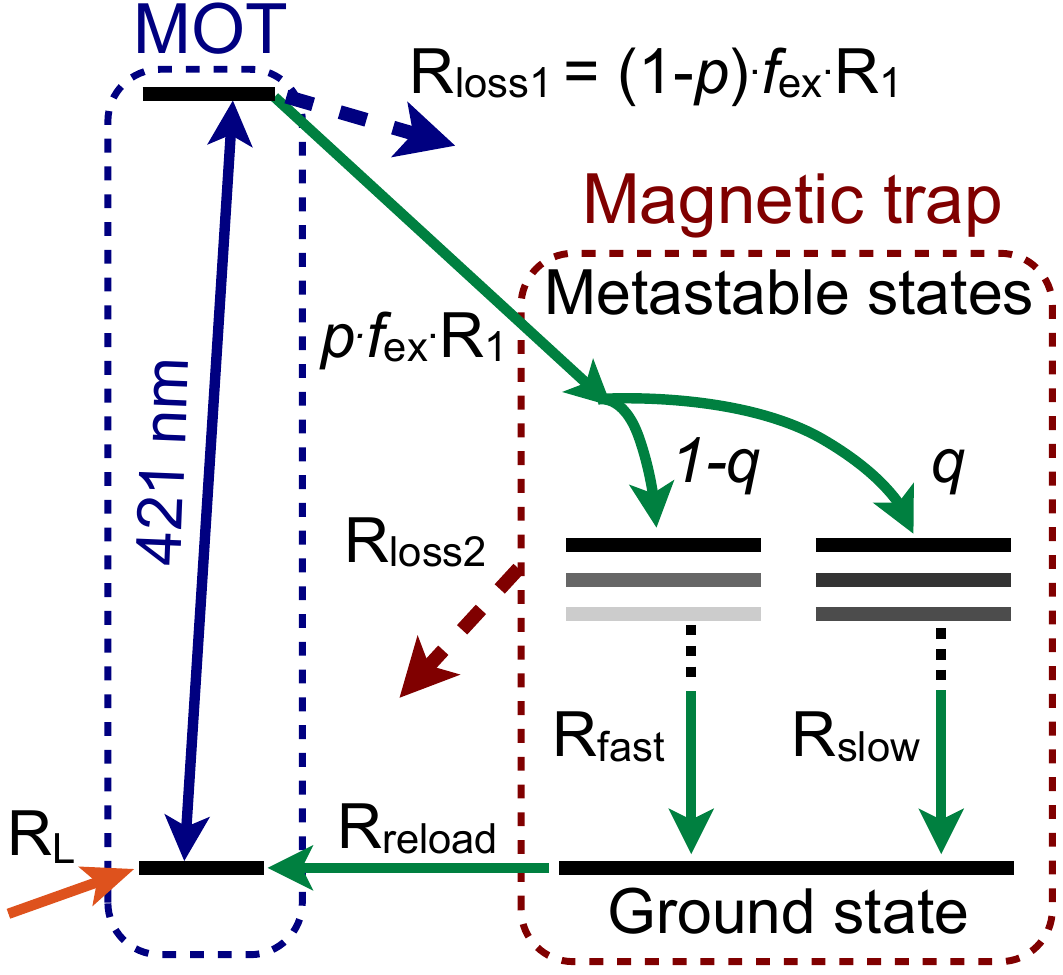}
\caption{\label{MOTscheme}(color online).   Dy MOT recycling and continuously loaded MT schematic.}
\end{figure}

As mentioned in Sec.~\ref{lasercooling}, the Dy MOT functions without any repumpers due to a population recycling mechanism based on its extraordinarily large magnetic moment.  Figure~\ref{MOTscheme} depicts a semiclassical rate equation model describing this recycling process.  Population is lost from the excited state of the MOT at rate $f_\text{ex}R_{1}$---which increases for larger excited state fractions $f_\text{ex}$---and decays to one of the (possibly many) metastable states (see Fig.~\ref{fig:dy_levels}).  A fraction of atoms $1-p$ are either in the $m=0$ Zeeman substate or in a strong magnetic field seeking $m$ state and never become confined in the MT.  The remaining atoms decay either through a fast decay channel at rate $R_\text{fast}$ and fractional population $1-q$ or through a slow channel at rate $R_\text{slow}$.  Atoms whose electronic population reaches the ground state are reloaded into the MOT at rate $R_\text{reload}$, which depends on the total MOT beams' saturation parameter, $s=\bar{I}/(1+(2\Delta/\Gamma)^2)$, where $\bar{I}=I/I_\text{s}$ and $\Delta$ is the detuning from resonance.  There must be some loss $R_\text{loss2}$ from the magnetic trap as the population cascades, but as shown in Ref.~\cite{Lu2010}, spin relaxation loss is $20\times$ slower than the timescales of the measured $R_\text{fast}$ and $R_\text{slow}$. 
\begin{figure}[t]
\includegraphics[width=0.4\textwidth]{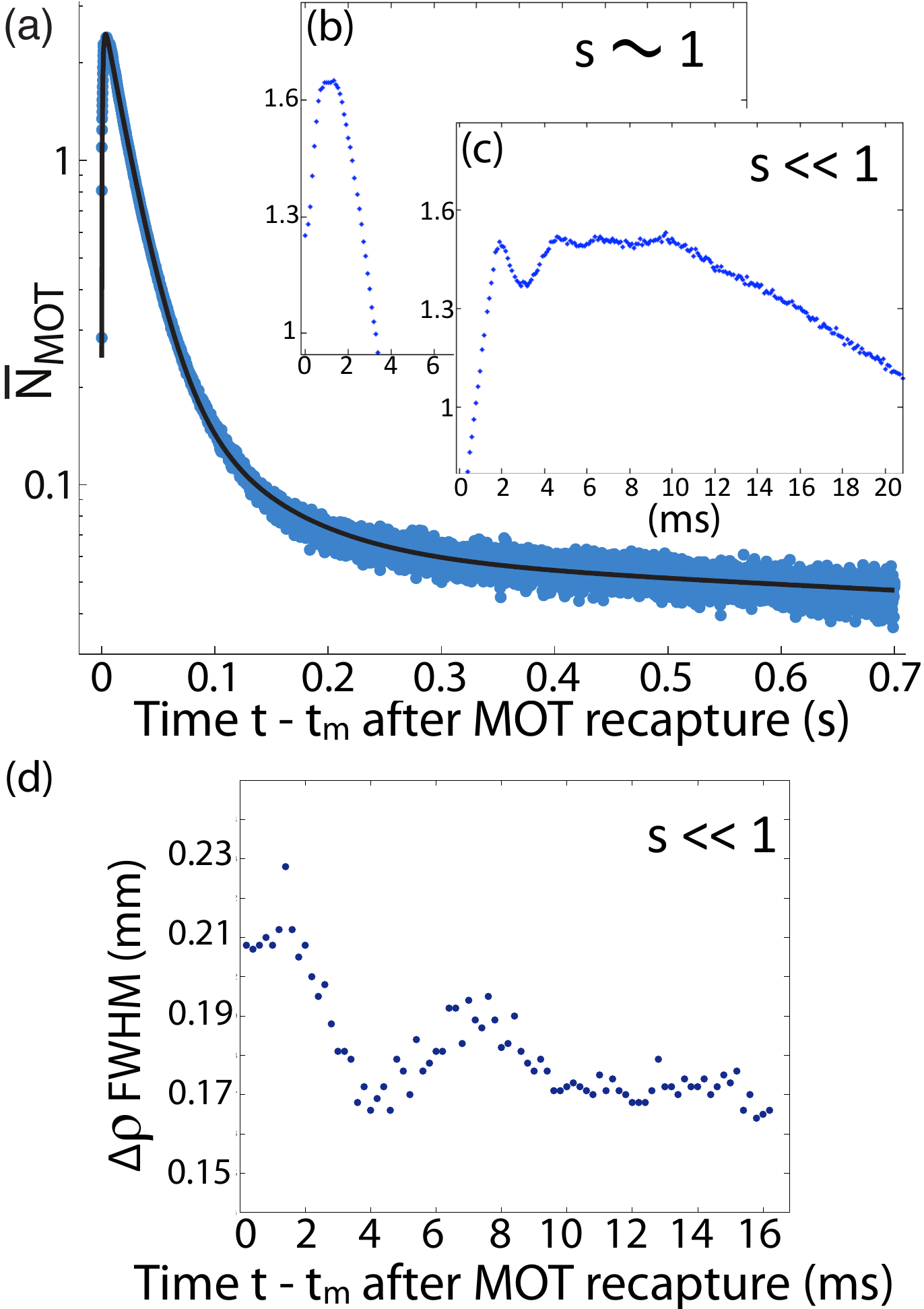}
\caption{\label{oscillations}(color online).  (a) Population ratio $\bar{N}_{\text{MOT}}$ of recaptured MOT to steady state MOT.  Black line is fit using Eqs.~\ref{MOTdecayeqns} to $\bar{N}_{\text{MOT}}$ with Zeeman slower and atomic beam off, and $t_m = 1$ s delay between steady state MOT and recapture. (b)-(c) Oscillations in fluorescence at the peak of the recaptured MOT population appear for MOT beam saturation parameters $s\ll1$.  (d) The Dy MOT operates in the mechanically underdamped regime, as shown by breathing mode oscillations, and the photon scattering rate seems to be correlated with cloud diameter.}
\end{figure}

The diagram in Fig.~\ref{MOTscheme} may be represented with the following rate equations: 
\bea\label{MOTdecayeqns}
\dot{N}_{\text{MOT}}&=&R_{\text{reload}}N_{\text{MT}}-f_{\text{ex}} R_1 N_{\text{MOT}}+R_{\text{L}}, \nonumber \\
\dot{N}_{\text{fast}}&=&(1-q)pf_{\text{ex}} R_1 N_{\text{MOT}}-R_{\text{fast}}N_{\text{fast}}, \nonumber \\
\dot{N}_{\text{slow}}&=&qpf_{\text{ex}} R_1 N_{\text{MOT}}-R_{\text{slow}}N_{\text{slow}}, \nonumber  \\
\dot{N}_{\text{MT}}&=&R_{\text{fast}}N_{\text{fast}}+R_{\text{slow}}N_{\text{slow}}-R_{\text{reload}}N_{\text{MT}},
\eea
where $N_{\text{MOT}}$, $N_{\text{fast}}$, $N_{\text{slow}}$, and $N_{\text{MT}}$ are the populations of the MOT, fast (slow) metastable state decay channel, and ground state MT, respectively, and $R_{\text{L}}$ is the loading rate from the Zeeman slower.  Fits of these rate equations to MOT population decay curves allow the determination of all free parameters for both the $^{164}$Dy boson and the $^{163}$Dy fermion:  \bea\label{MOTdecayResults}
\left[R^{163}_1, R^{163}_{\text{fast}}, R^{163}_{\text{slow}}\right]
 &=&[1170(20),  19(2),1.5(1)]\text{ s}^{-1},  \\
\left[R^{164}_1, R^{164}_{\text{fast}}, R^{164}_{\text{slow}}\right]
 &=& [1700(100),  29(1), 2.3(1)]\text{ s}^{-1}, \nonumber \\
 \left[p, q\right] &=& [0.82(1), 0.73(1)], \nonumber
\eea
where $p$ and $q$ for the two isotopes are equal within error.

\begin{figure}[t]
\includegraphics[width=0.49\textwidth]{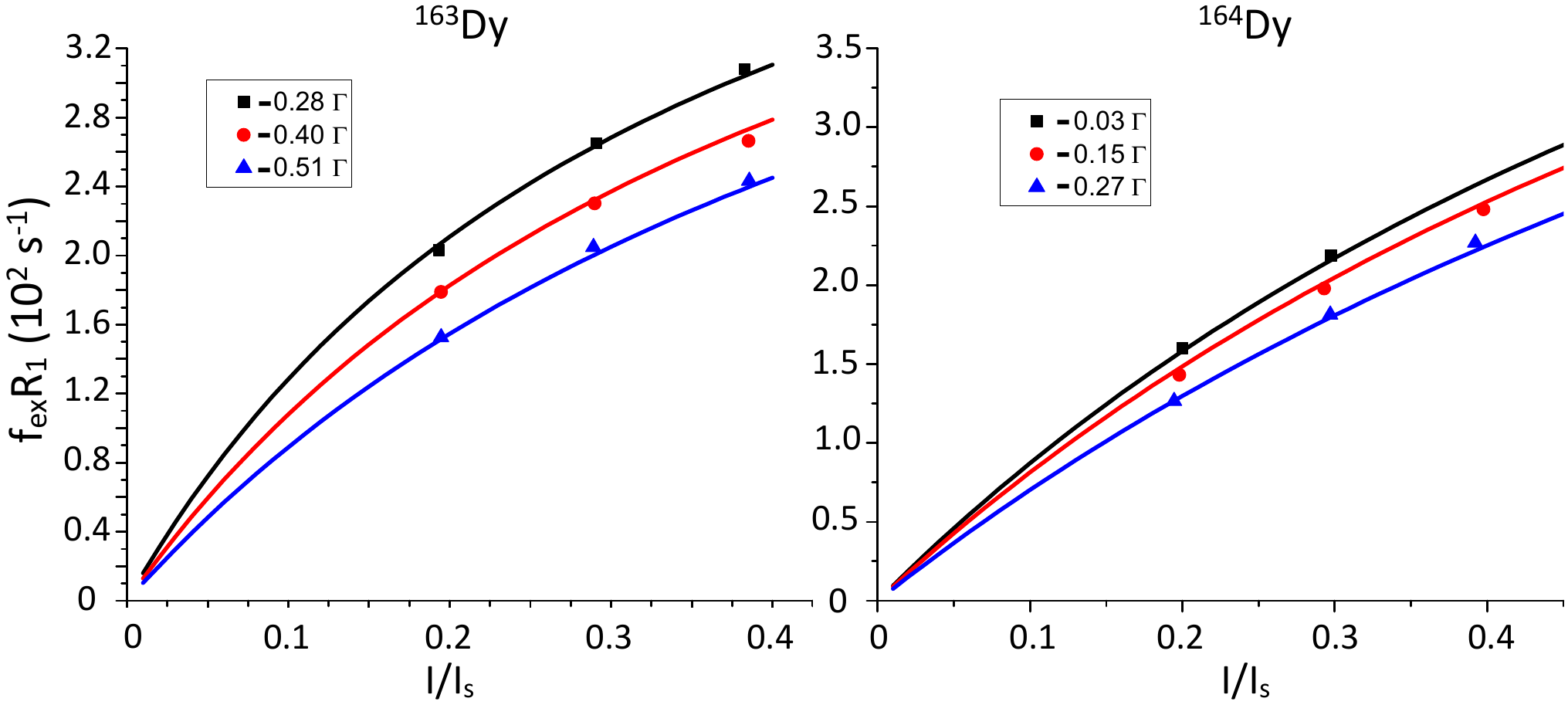}
\caption{\label{fig:fexR1}(color online).  Fits of Eq.~\ref{fexR1eqn} to $f_{ex}R_{1}$ versus $\bar{I}=I/I_{s}$ and $\Delta$ for both fermionic $^{163}$Dy and bosonic $^{164}$Dy.  A simultaneous fit to the nine (eight) data points for $^{163}$Dy ($^{164}$Dy) allow the extraction of $R_{1}$ from the product $f_{ex}R_{1}$ via Eq.~\ref{fexR1eqn}.}
\end{figure}
An example of the Dy MOT decay and fit are shown in Fig.~\ref{oscillations}(a).  To obtain these data, we collect MOT fluorescence using a 200-mm 2'' aspherical lens and focus it through a pinhole, using a 60-mm 2'' aspherical lens.  The light is then collected on an amplified, high-bandwidth PIN photodetector.  We load the MOT for 2--3 s to reach steady-state population before extinguishing, at $t_{m}=0$, the MOT beams, the Zeeman slower beam, and the atomic beam (via the in-vacuum shutter).  Note, the magnetic quadrupole field remains on throughout this experiment, providing confinement for weak-field seeking metastable and ground state atoms.  We wait for $t_{m}=1$ s before recapturing the MOT by turning on the MOT beams, but we leave the Zeeman slower beam and atom beam off to ensure there is no external loading of the MOT during the measurement, i.e., $R_{L}=0$.   Shorter $t_{m}$ times bias the results due to ill-defined initial conditions:  we fit the data assuming nearly all population has decayed to the ground state, allowing us to set $N_{\text{MOT}} = N_{\text{fast}} = N_{\text{slow}}= 0$ at $t-t_{m}=0$, while $N_{\text{MT}}$ is left free to vary and is normalized to the steady state MOT fluorescence signal.  Fluorescence traces as in Fig.~\ref{oscillations}(a) are averaged over 16 sequential runs on a digital oscilloscope.

We noticed an oscillation of the recaptured MOT fluorescence near peak signal when operating at small saturation parameter $s\ll1$.  The pinhole does not restrict numerical aperture so severely that the modulation is simply due to a loss of photons when the cloud is large.  This oscillation disappears when the intensity of the MOT light is increased and/or the detuning of the beams decreased to the point that $s\sim1$ [compare Figs.~\ref{oscillations}(b) and (c)].  Data corrupted with oscillations could not be fit using Eqs.~\ref{MOTdecayeqns}, so all the data used to extract $R_{1}$, $R_{\text{fast}}$, $R_{\text{slow}}$, $p$, and $q$ were taken near $s\sim1$.  

A possible explanation of these fluorescence oscillations lies in the motional dynamics of the Dy MOT.  Most MOTs, e.g., of Rb, are solidly in the overdamped regime, meaning they have a ratio $\alpha=\Gamma_{\text{MOT}}/\omega_{\text{MOT}}>1$ of the MOT's optical damping rate $\Gamma_{\text{MOT}}=\beta/2m$ to the trap oscillation frequency $\omega_{\text{MOT}}=\sqrt{\kappa/m}$.   In a 1D treatment, the damping coefficient $\beta$ is \be\label{beta} 
\beta = \frac{8\hbar k^{2}|\Delta| \bar{I}}{\Gamma(1+\bar{I}+(2\Delta/\Gamma)^{2})^{2}},
\ee where $k$ is the wavenumber of the MOT light, mass is $m$, and the spring constant is $\kappa = \mu'\nabla B \beta/\hbar k$~\cite{MetcalfBook99}.  In this latter expression, $\mu'\equiv(g_{e}m_{e}-g_{g}m_{g})\mu_{B}$, where $g_{i}$ and $m_{i}$ are the $g$ factors~\footnote{For Dy, $g_{g}=1.24$ and $g_{e}=1.22$.} and Zeeman substates of the ground and excited levels.  For a typical Rb MOT operated to maximize trap population, $\alpha_{\text{Rb}}\agt7$.  For the Er MOT with  the parameters used in Ref.~\cite{Mcclelland:2006}, $\alpha_{\text{Er}}=1.3$ and oscillations where not reported~\cite{Andy2009}.  However, the Dy MOT, when operated in the regime that maximizes trapped atom population (see Sec.~\ref{stripes}), is just in the underdamped regime $\alpha_{\text{Dy}}=0.8$.  The origin of this lower $\alpha$ lies in the combination of the larger (smaller) $m$, $k$, $\Gamma$, $\mu'$ and $\nabla B$  ($\bar{I}$ and $\Delta/\Gamma$) parameters in the highly magnetic Dy MOT versus typical Rb MOTs.  

Indeed, $\alpha_{\text{Dy}}=0.8$ corresponds to an oscillation period of 4 ms, and a measurement of the breathing mode of the Dy MOT, shown in Fig.~\ref{oscillations}(d), shows a damped oscillation with a period of $\sim$$6$ ms after the magnetic quadrupole field gradient is decreased by a factor of 2 at $t_{m}=0$. We find that the breathing mode period scales $\propto\sqrt{\nabla B}$.  The similar periods support the notion that fluorescence is modulated by the motion of the atoms during recapture.  (Atoms confined in the magnetic trap, from which the MOT is recaptured, possess a different spatial profile from those in the MOT.)  The breathing mode should have a period half that of the trap oscillation frequency, indicating that the actual $\alpha_\text{Dy}$ is just slightly below unity; a 3D numerical MOT calculation incorporating magnetostatic forces on the highly magnetic Dy could better estimate $\alpha$ for these system parameters.   We conjecture that optical pumping, Zeeman shifts, and spin polarization near the cloud edge---as well as modulation of optical density---could change the effective photon scattering rate as the cloud expands and contracts.  To avoid complications caused by these motional dynamics, we operate the MOT in the overdamped regime of $s\sim 1$ for the decay rate measurements.

To extract MOT decay rates $R_{1}$ from the product $f_{ex}R_{1}$---as well as to obtain better error estimates of the other model parameters---we repeat the measurement in Fig.~\ref{oscillations}(a) for several combinations of $\bar{I}$ and $\Delta$.  MOTs of bosonic $^{164}$Dy and fermionic $^{163}$Dy are studied in a similar manner to investigate differences in decay dynamics due to hyperfine structure.  Figure~\ref{fig:fexR1} shows the set of data simultaneously fit to the function \be \label{fexR1eqn}
R(I,\Delta)=R_1\bar{I}/(2+2\bar{I}+2(2\Delta/\Gamma)^2),
\ee
where $R = f_{\text{ex}} R_1$.  Using this method to determine $R_{1}$, we arrive at the rates in Eq.~(\ref{MOTdecayResults}) for these two Dy isotopes.   

Detailed simulations of the decay channels still seem beyond the reach of tractable calculations, but such calculations might be able to employ the following simplification for modeling the fast decay channel.  The shortest possible decay channel (the fast channel), must involve at minimum two metastable levels due to the need to switch parity from odd (421-nm exited state) to even to odd before decaying back to the even parity ground state.  Recent calculations reported in Ref.~\cite{Flambaum2010} indicate that the lifetimes for the 1001-nm and 1322-nm levels are $2\pi\cdot[53, 23]$ s$^{-1}$, respectively, which indicates that rapid decay to these levels followed by a delay given by their lifetime is a likely candidate for the 20--30 s$^{-1}$ fast decay channel.  For the slow decay channel, it is possible that the population becomes shelved in a longer-lived state that decays to ground or an intermediate state via a non-electric dipole (E1) allowed transition.  Small-energy-difference decays could also contribute to the slow channel. 

The large $q$ parameter---73\% of the atoms decay through the slow versus fast channel---serves a very useful purpose in providing a continuously loaded magnetic reservoir of metastable atoms for the MOT.  Indeed, data presented in Ref.~\cite{Lu2010} show that 2.5$\times$ ($^{164}$Dy) to 3.5$\times$ ($^{163}$Dy) more atoms are in the metastable MT plus MOT than in the visible MOT alone, resulting in a maximum number of laser cooled and trapped atoms nearing 5$\times$10$^{8}$ for the Dy system.  This is reminiscent of the continuously loaded MT in the Cr system~\cite{Pfau01MT} that provides sufficiently large samples, $10^{8}$ atoms, to enable Bose-condensation of Cr~\cite{Pfau05CrBEC}.  

\subsection{MOT population versus isotope}\label{shelving}
\begin{figure}[t]
\includegraphics[width=0.49\textwidth]{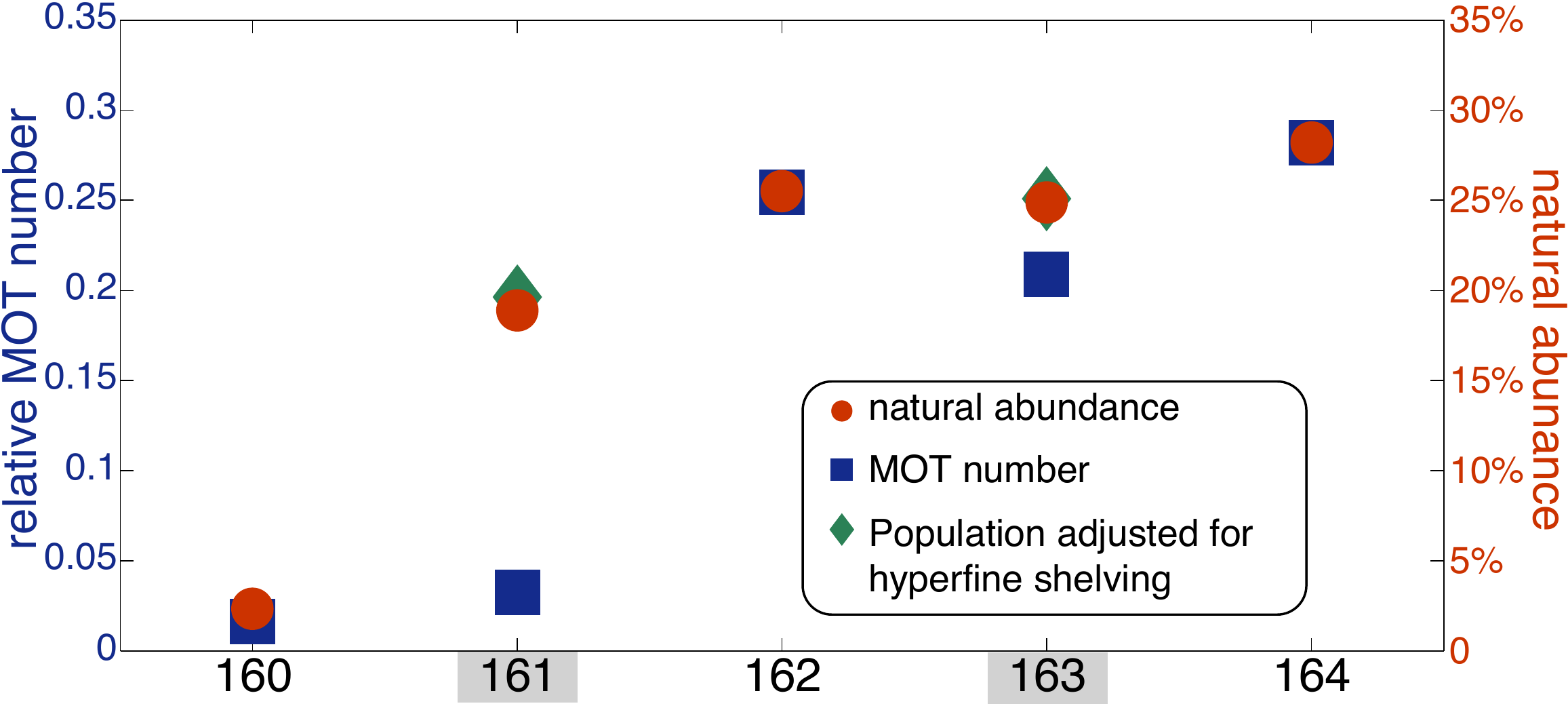}
\caption{\label{isotopes}(color online).  Comparison of relative MOT population to natural isotope abundance for the five highest abundance isotopes of Dy.  Relative population is computed as the fraction of an isotope's MOT population out of the sum total MOT population for all the isotopes.  The fermions, $^{163}$Dy and $^{161}$Dy are highlighted in grey.  The green diamonds are the MOT populations for the fermions when multiplied by $6/5$ for $^{163}$Dy and 6 for $^{161}$Dy to account for poor optical pumping of the six hyperfine states during laser cooling.  See text for details.}
\end{figure}

Figure~\ref{isotopes} plots the maximum MOT population obtained for each stable isotope (except for low 0.1\% abundance $^{158}$Dy) along with the natural abundance of the isotope.  To facilitate comparisons to natural abundance, each MOT population is normalized by the sum of all the MOT populations of the isotopes (for the fermionic isotopes, the MOT population is adjusted for hyperfine shelving; see below).   The bosonic MOTs have relative populations very close to their proportion of the natural abundance.  However, the fermionic isotopes $^{163}$Dy and $^{161}$Dy have populations that are 84.0\% and 17.3\% their natural abundances, respectively.  We try, in the following, to argue why repumper-less fermionic MOTs form and are observed with these population ratios.

Unlike the bosons, the two fermions have non-zero nuclear spin $I=5/2$, and the opposite sign of the nuclear magnetic moment between the two isotopes results in oppositely ordered hyperfine levels with respect to $F$ versus energy [see Fig.~\ref{fig:dy_levels} (b) and (c)].  With a total electronic angular momentum of $J=8$, the isotopes have six hyperfine states, $F=11/2$ to $21/2$ and $F'=13/2$ to $23/2$, in their ground and 421-nm excited states, respectively.  Without hyperfine repumpers (a repumper is necessary for Rb and Cs MOTs), one would expect that no MOT could form on the $F=21/2\rightarrow F'=23/2$ cycling transition due to rapid decay to $F<21/2$ states.  Alkalis such as Rb and Cs, however, have smaller hyperfine splittings ($<$270 MHz) between their highest $F'$ excited states than does $^{163}$Dy, whose splitting is 2.11 GHz.  However, such an explanation---that scattering to the lower $F$ hyperfine state is slower due to the larger detuning---ignores the 6$\times$ shorter lifetime of the Dy excited state.  Together this implies a depumping rate similar to Rb and Cs.  The crucial difference, however, lies in the fact that the hyperfine splittings in the ground and excited states of Dy are nearly matched, whereas for Rb and Cs, they are different by more than 25$\times$.  Thus for Dy, the trapping light near-resonant with the $F=21/2\rightarrow F'=23/2$ cycling transition also serves as an efficient repumper for the  $F=19/2\rightarrow F'=21/2$ transition, thereby preventing complete population shelving into dark states. 

A full modeling of the intra-hyperfine manifold population transfer as the atoms progress through the transverse cooling, Zeeman slower, and MOT stages is beyond the scope of this paper, but we can make a simplified estimate of the $F=21/2$ state decay and repumping rates in each of the stages.  The atoms spend roughly 20~$\mu$s in the transverse cooling beams, scatter $\sim$$8\times10^{4}$ photons from the $F'=23/2$ state in the Zeeman slower, and spend, on average, 25 ms in the MOT scattering photons from the 421-nm excited state before decaying to metastable states.   Accounting for laser intensities (power broadening) and detunings, this implies that for $^{163}$Dy atoms, $<$1 photon is off-resonantly scattered on the $F'=21/2$ state (which can decay to $F=19/2$) in the transverse cooling stage, while nearly 100 and 50 are scattered in the Zeeman slower and MOT stages, respectively.  However, the repumping rates in the Zeeman slower and MOT stages back to the $F'=21/2$ state from $F=19/2$ are only a factor of two smaller than the cycling transition scattering rate and are 570$\times$ and 1300$\times$ faster than the decay rates to $F=19/2$, respectively:  in very little time the population is repumped back into the cycling transition.  No secondary laser is necessary to repump the $^{163}$Dy system.  

The situation is different in the $^{161}$Dy system due to the hyperfine structure inversion and the nearer detuning of hyperfine levels at the larger $F$ and $F'$ end of the spectrum.  In the transverse cooling, Zeeman slowing, and MOT stages, approximately 10, $3\times10^{3}$, and $2\times10^{3}$ photons are off-resonantly scattered to the $F'=21/2$ state, respectively; the $^{161}$Dy system off-resonantly scatters more than in the $^{163}$Dy system.  Nevertheless, the repumping rates are within a factor of two of the depumping decay rates, and atoms are repumped nearly as fast as they are depumped from the cycling transition.  Again, a repumper-less MOT is able to be formed with $^{161}$Dy despite its multitude of hyperfine levels.   Reference~\cite{Mcclelland:2006} reported a fermionic $^{167}$Er MOT ($I=7/2$), which we suspect works in a similar manner.

This leaves open the question why we observe as much as 84\% of the natural abundance of $^{163}$Dy, while as little as 17\% of $^{161}$Dy.  The hyperfine population in Dy arriving from the high-temperature oven is likely to be distributed with a bias toward the higher-multiplicity large-$F$ states.  This may contribute to the large MOT population in $^{163}$Dy, but $^{161}$Dy's much lower population may be due to the competing effect of less-efficient optical pumping and repumping to the $F=21/2$ state.

Optical pumping of population into the $F=21/2$ state must occur before the Zeeman slower detunes the non-$F=21/2$ state atoms away from resonance.  Indeed, we have already seen that the 421-nm laser tuned to the $F=21/2\rightarrow F'=23/2$ transition can pump atoms out of the $F=19/2$ state on relatively short time scales, and we now estimate all the relative $F\rightarrow F'=F+1$ effective scattering rates to examine the repumping efficiency from the $F\leq19/2$ states:
\bea\label{ratios}
\left[F^{163}_{21/2}:F^{163}_{19/2}:F^{163}_{17/2}:F^{163}_{15/2}:F^{163}_{13/2}:F^{163}_{11/2}\right] = \ \ \ \ \ \ && \\ 
  \left[1000:500:70:20:6:3\right]. &&\nonumber \\
\left[F^{161}_{21/2}:F^{161}_{19/2}:F^{161}_{17/2}:F^{161}_{15/2}:F^{161}_{13/2}:F^{161}_{11/2}\right]  = \ \ \ \ \ \ &&\nonumber \\
\left[1000:20:7:5:5:7\right]&&. \nonumber 
\eea
These ratios are proportional to the number of photons scattered from the $F=21/2\rightarrow F'=23/2$ cycling transition, denoted $F^{163,161}_{21/2}$ and normalized to  $F^{163}_{21/2}=F^{161}_{21/2}=1000$, during the time the atoms transit the transverse cooling stage and first 5 cm of the Zeeman slower~\footnote{This is roughly the portion of the Zeeman slower in which the fast oven beam is in resonance with the slowing laser.}.  $F^{163}_{11/2}=3$ for the $F=11/2\rightarrow F'=13/2$ transition corresponds to roughly 20 scattered photons in our experiment, assuming that the population is already optically pumped to the $m_{g}=F$ state~\footnote{The transition strengths are nearly equal for the five stretched-state $\sigma_{+}$ transitions.}.  This is merely an upper bound to the number of photons a non-polarized sample would scatter since $m_{g}<F\rightarrow m_{e}< F+1$ transitions have smaller transition strengths than the cycling transition $m_{g}=F\rightarrow m_{e}= F+1$.
\begin{figure}[t]
\includegraphics[width=0.49\textwidth]{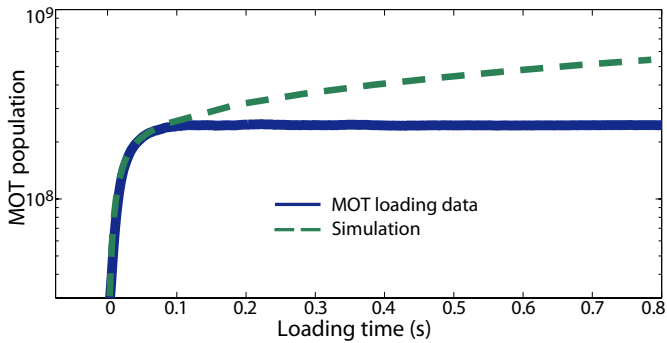}
\caption{\label{MOTLoad}(color online).  Typical MOT loading data as a function of time (blue, solid curve).  The trapped atom population is detected by a photodetector recording MOT fluorescence.  The Zeeman slower laser and atom beam shutters open at $t=0$.  The green (long-dash) curve is a simulation using Eqs.~\ref{MOTdecayeqns} and~\ref{MOTdecayResults}.}
\end{figure}

A striking difference between the isotopes is evident. The single 421-nm $F=21/2\rightarrow F'=23/2$ laser scatters many more photons from $F<21/2$ states in $^{163}$Dy than in $^{161}$Dy.  This leads to much more efficient optical pumping to the $F=21/2$ state in $^{163}$Dy.  We conjecture that this is why 84\% of the $^{163}$Dy are trapped but only 17\% of the $^{161}$Dy.  Coincidentally, these percentages are close---within experimental error---to $5/6$ and $1/6$, respectively.  It is tempting---but probably an oversimplification---to conclude that for $^{163}$Dy, all levels but the lowest $F$ state is optically pumped to $F=21/2$.  The case of $^{161}$Dy is certainly more complicated:  while it is likely that $F\leq17/2$ states are never efficiently pumped to $F=19/2$ (let alone $F=21/2$), the initially larger $F$ population from the oven should lead to larger populations than 17\%.   A possible explanation lies in the inversion of the hyperfine state energy hierarchy between the two isotopes.  This causes Doppler shifts in atoms not optimally decelerated in the Zeeman slower to move closer to (further from) resonance with the $F<21/2$ levels in $^{163}$Dy ($^{161}$Dy).  This contributes to a more efficient slowing---and therefore more efficient MOT loading---of $^{163}$Dy versus $^{161}$Dy atoms.  

A thorough optical pumping simulation including all the hyperfine levels and the actual experimental parameters and geometry could better address these questions.  A repumping laser system is under construction to identify into which $F<21/2$ states population accumulate, and we suggest that only one repumper on the $F=11/2\rightarrow F'=13/2$ transition for $^{163}$Dy would be necessary to capture the full fraction of this isotope's natural abundance in the MOT.  Reference~\cite{Leefer:2010} presents other repumping schemes.

\subsection{MOT loading}
Figure~\ref{MOTLoad} shows a typical MOT population loading curve after the Zeeman slower laser and atomic beam are unblocked at $t=0$.  The $^{164}$Dy MOT population---proportional to the MOT florescence (atoms in the metastable MT remain dark)---rises to a steady state of $2.5\times10^{8}$ within 50 ms (blue line).  A simulation of Eqs.~\ref{MOTdecayeqns} predicts, however, that within 5 s the population should reach a steady state 5$\times$ larger (green dashed line).  The Dy MOT population limit could arise from light-induced two-body collisions as in the Cr MOT~\cite{Bradley2000}.  Suppression of the $R_{\text{reload}}$ term in Eqs.~\ref{MOTdecayeqns} mimics the experimental data for a loading rate of $R_{\text{L}}\approx10^{10}$ s$^{-1}$.  Reproduction of the loading data requires a small $R_{\text{reload}}$ in simulations, while fits to decay data such as in Fig.~\ref{oscillations}(a) require $R_{\text{reload}}$ to be of the order 10$^{3}$ s$^{-1}$.  A main difference between the MOT loading and MOT decay experiments is the presence of the 1 W Zeeman slowing beam, which could enhance light-induced losses of Dy and reduce $R_{\text{reload}}$.  Another possibility involves 2-photon ionization loss while the atoms are in the metastable states (twice the cooling light energy is close to the ionization potential)~\cite{Neefer2010PC}.

\subsection{Comparison to Er MOT decay dynamics} 
As discussed earlier, the Er MOT did not exhibit underdamped oscillations in MOT decay florescence, which we believe is due to its smaller magnetic moment.  Other differences between the Dy and Er MOT decay dynamics are presented in this section.

The analysis of the Er MOT decay dynamics~\cite{Mcclelland:2006} accounted for loss only via an $R_{\text{lossMT}}$ term (i.e., $p\equiv1$) and with only one decay channel through the metastable states ($q\equiv0$)~\footnote{Including an $R_{\text{lossMT}}$ loss term is equivalent to allowing $p$ to be non-unity, but the latter can be more physically motivated.}.  Allowing $q$ to be nonzero (thus introducing two decay channels), provides a much better fit to the Dy MOT decay data.   The Er apparatus had a 1000$\times$ worse vacuum pressure, which might have obscured the long-time tail of the MOT decay and prevented a measurement of a second metastable state decay channel.  

Perhaps coincidentally, the bosons $^{164}$Dy and $^{168}$Er have the same, within experimental error, rate of decay to metastable states $\sim$1700 s$^{-1}$ and branching ratios.  Interestingly, the fermion $^{163}$Dy, which possesses hyperfine structure, has a significantly smaller decay rate 1170 s$^{-1}$ as well as shorter metastable decay rates compared to $^{164}$Dy.   While it is not clear why all the  $^{164}$Dy decay rates are $\sim$1.5$\times$ larger than those for $^{163}$Dy, the existence of hyperfine structure in $^{163}$Dy likely plays a role.  

As shown in Ref.~\cite{Lu2010}, $^{163}$Dy's smaller decay rates translate into larger steady state metastable MT populations, which help to enhance its total trapped population.  The large $q$ and small $R_{\text{slow}}$ that provide population enhancement in the metastable MT was also seen in MOT-loaded magnetic traps of $^{168}$Er, albeit to a lesser extent due to poor lifetimes from a high vacuum~\cite{Berglund:2007}.  The majority of Dy atoms decay through a slower metastable channel (2.3 (1.5) s$^{-1}$ versus 4.5 s$^{-1}$ for $^{168}$Er).  However, 27\%  decay through the fast channel at rates 6$\times$ (4$\times$) larger than $R_\text{slow}$ in $^{164}$Dy ($^{163}$Dy).  From simulations of Eqs.~1 in Ref.~\cite{Mcclelland:2006} and Eqs.~\ref{MOTdecayeqns} here, one can see that the measured rates imply that the Er metastable MT reservoir could hold several times more atoms as Dy's; however, the severity of Er's magnetic trap inelastic loss rates are not yet known.  Dy collisions limited the metastable trap population to 80\%~\cite{Lu2010} of the maximum achievable given by the rates in Eqs.~\ref{MOTdecayResults}.  

\subsection{MOT population and temperature}\label{stripes}

\begin{figure}[t]
\includegraphics[width=0.49\textwidth]{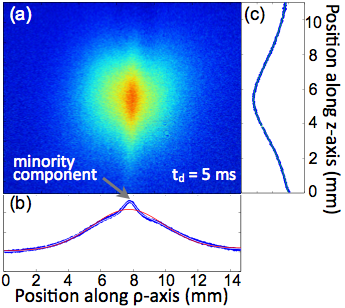}
\caption{\label{stripe3ms}(color online).  (a) Time-of-flight absorption image of $^{164}$Dy MOT at $t_{d}=3$ ms; MOT fields are extinguished at $t_{d}=0$.  Intensity integrations along (b) $z$ and (c) $\rho$-directions.  Grey arrow points to region of minority component for this MOT.  (b) Double-Gaussian fit (white line) to MOT expansion (the red line is a single-Gaussian which results in a poor fit).  The majority component (hot outer cloud) is defined as the atoms contributing to the broader Gaussian, while atoms in the narrower Gaussian comprise the minority component. (c)  The anisotropically cooled stripe hampers majority component temperature measurements in the $z$-direction. The origin and temperature characteristics of the minority component are studied in Ref.~\protect\cite{Youn2010b}.}
\end{figure}
\begin{figure}[t]
\includegraphics[width=0.49\textwidth]{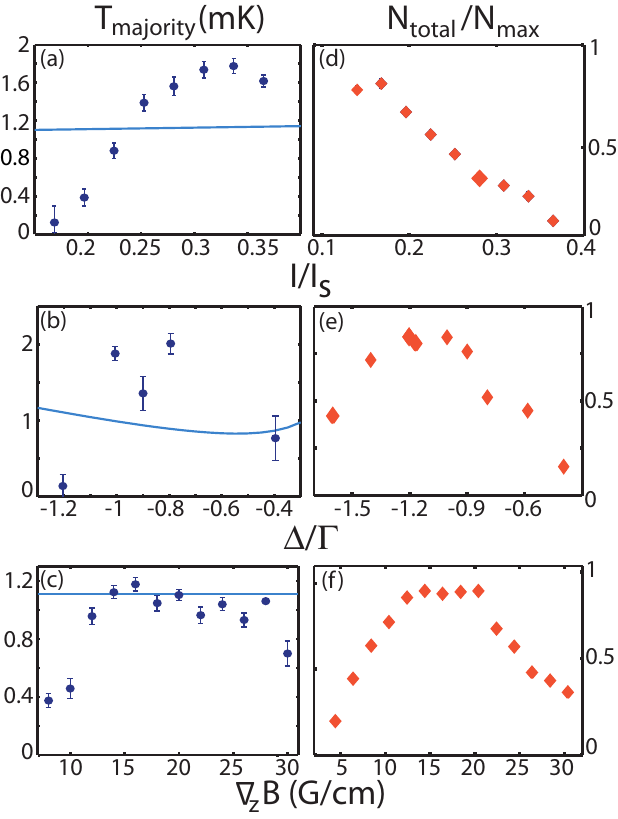}
\caption{\label{Ma}(color online).  (a--c) Temperature characterization of the majority component of the Dy MOT versus intensity $\bar{I}=I/I_{s}$, detuning $\Delta/\Gamma$, and MOT magnetic field gradient $\nabla_{z}$B ($\nabla_{\rho}\text{B}\approx\nabla_{z}\text{B}/2$).  Data are the temperatures of the MOT in the $\rho$ direction (quadrupole plane of symmetry, gravity points in the $-z$ direction).  Light blue curves are plots of Eq.~\ref{MajorTN} for the following parameters:  (a,d) $\Delta/\Gamma = -1.2$, $\nabla_{z}\text{B}=20$ G/cm; (b,e) $\bar{I} = 0.17$, $\nabla_{z}\text{B}=20$ G/cm; (c,f) $\bar{I} = 0.20$, $\Delta/\Gamma = -1$.  (d--f) Total number of atoms (major plus minor component) versus $\bar{I}$, $\Delta/\Gamma$, and $\nabla_{z}$B.  The typical visible MOT maximum population is $N_{\text{max}}=2\times10^{8}$.  Maximum density of $\sim$$10^{10}$ cm$^{-3}$ occurs for: $\bar{I}=0.33$ in (a,d); $\Delta/\Gamma=-1$ in (b,e); and $\nabla_{z}\text{B}=20$ G/cm in (c,f).}
\end{figure}

As reported in Ref.~\cite{Lu2010}, when the power between the three sets of MOT beams are mismatched, the Dy MOT can exhibit a sub-Doppler cooled core, which depending on the details of the power balancing, can be anisotropic in its temperature distribution.  However, the number of atoms in this ultracold core is typically less than 10\% of the total MOT population.  We leave the study of this minority component to Ref.~\cite{Youn2010b}, and instead present here the temperature and population of the majority component of the Dy MOT as a function of MOT beam intensity $\bar{I}$, detuning $\Delta/\Gamma$, and quadrupole gradient $\nabla B_{z}$ (see Fig.~\ref{stripe3ms}).  Data presented here are typical of the majority component in all Dy MOTs.

To distinguish the hotter majority from the colder minority component---when it exists---we fit a double-temperature distribution to the time-of-flight expansions of the MOT (Fig.~\ref{stripe3ms}(b) and Fig. 4 in Ref.~\cite{Lu2010}).  Figure~\ref{Ma} shows a compilation of temperature and population data of the majority component.  In both Figs.~\ref{stripe3ms} and~\ref{Ma}, the data were taken for a ratio of MOT powers in the $z$ versus $\rho$ directions of $I_{z}/I_{\rho}\approx1$ and $\bar{I}>0.2$ ($I = I_{z}+2I_{\rho}$).  This results in a MOT with a vertically oriented, anisotropically sub-Doppler cooled core.   We choose to discuss the majority component data with the MOT beams set to $I_{z}/I_{\rho}\leq1$ and $\bar{I}>0.2$ since it is in this regime that we obtain the most populous MOTs.

The major component temperature $T_{\text{majority}}$ data in the $\rho$-direction are shown along with plots of Eq.~\ref{MajorTN}, the temperature $T_{\text{D}}$ from simplified Doppler cooling theory~\cite{MetcalfBook99}:
\be\label{MajorTN}
T_{\text{D}}=\frac{\hbar\Gamma}{4k_{B}}\frac{1+\bar{I}+(2\Delta/\Gamma)^{2}}{2|\Delta|/\Gamma}.
\ee
The Doppler cooling limit for Dy on the 421-nm transition is 770 $\mu$K, and for the $\bar{I}$ and $\Delta/\Gamma$'s typically used in the experiment, the Doppler cooling temperature is $\sim$1 mK.  The prolate MOT $T_{\text{majority}}$ data  are close to that predicted by Doppler cooling theory for large $s$ (large $\bar{I}$, small $\Delta/\Gamma$) and high gradients.  However, $T_{\text{majority}}$ of the atoms at small $s$ and small gradient is below the Doppler limit.  

The low gradient and low saturation data are consistent with the several hundred $\mu$K temperatures of the Er MOT~\cite{Berglund:2007}.  One-dimensional numerical sub-Doppler cooling simulations presented in Ref.~\cite{Berglund:2007} indicate that population-wide MOT sub-Doppler cooling arises from the near-equal {Land\'{e}} $g$ factors in Er's ground and excited states $\Delta g_{\text{Er}}=0.004$, which means that the MOT magnetic field Zeeman shifts the ground and excited $m_{J}$ levels by nearly equal energies.  Spherically symmetric intra-MOT sub-Doppler cooling has been observed in the cores of MOTs of more commonly used atoms (e.g., Rb~\cite{Jhe:2004}), but unlike Er, the majority of atoms are at hotter Doppler temperatures.  The Dy MOT---which also possesses a near-degeneracy of $g$ factors on the 421-nm line~\footnote{$\delta g_{\text{Dy}}$ = 0.022 ($\delta g_{\text{Dy}}/g_{\text{Dy}}=1.7$\%), which is 5.5$\times$ larger than Er's on its MOT transition, but 7.7$\times$ less than Rb's.}---also exhibits population-wide MOT sub-Doppler cooling like the Er MOT, which is confirmed in 1D numerical simulations~\cite{Andy2009,Lu2010}.  Unlike Er, however, the Dy MOT population assumes a double-temperature distribution at large $s$ and $\nabla_{z} B$.

In contrast to the Er MOT, we have also observed anisotropic sub-Doppler cooling to much lower temperatures---down to 10 $\mu$K---in the central cores of the Dy MOTs.  This phenomenon of ultracold stripes, shown in Fig.~\ref{stripe3ms}, disappears at low MOT beam intensities---coincident with the appearance of sub-Doppler cooling in the majority component of the Dy MOT.  The minority stripe components are typically 1-10\% the population of the entire MOT, and as such are no more populous than $10^{6}$ to $10^{7}$ atoms (see Fig.~\ref{stripe3ms}).  The Er MOT presented in Ref.~\cite{Berglund:2007} contained no more than $2\times10^{5}$ atoms, and the largest Er MOT to date contained $1.6\times10^{6}$ atoms~\cite{Mcclelland:2006}.  We conjecture that the Er population was limited by low Zeeman slower power, which was $<$100 mW~\cite{Andy2009}.  More populous Er MOTs might also exhibit double temperature distributions when larger MOT beam power is employed.  The origin of this anisotropically sub-Doppler cooled minority component located in the Dy MOT core is explored in detail in Ref.~\cite{Youn2010b}.

\section{Conclusion}

We presented a detailed description of the creation of large-population open-shell lanthanide MOTs with the repumper-less technique.  The demonstration that the repumper-less MOT technique works for Dy in a manner similar to Er (but with important differences) lends support to endeavors aimed at forming MOTs of other interesting highly magnetic lanthanides such as holmium~\cite{Saffman:2008}.  We also presented an in-depth characterization of the Dy MOT population, temperature, loading rate, isotope trapping efficiency, mechanical dynamics, and metastable state recycling dynamics.  Future work will attempt to narrow-line laser cool Dy on the 741-nm transition in a manner similar to that demonstrated for Er~\cite{Berglund:2008}.  

We note that Dy MOTs made in a similar manner to alkaline-earth MOTs~\cite{ErTransitions05,Grimm09}---i.e., using the 421-nm light for Zeeman slowing (and transverse cooling) but creating the MOT on the closed $140$ kHz-wide 598-nm transition~\cite{Martin:1978}---will likely be possible and would provide lower initial temperatures.  However, for maximizing phase-space density, it is not yet clear whether a lower initial temperature outweighs the loss of the metastable MT reservoir and associated population gains.  It remains to be seen whether such MOTs could also trap fermionic isotopes without additional repumpers.

\begin{acknowledgments}
We thank A{.} J{.}~Berglund, J{.} J{.}~McClelland, E{.} Fradkin, N{.} Leefer, and D{.} Budker for helpful discussions.  We acknowledge support from the NSF (PHY08-47469), AFOSR (FA9550-09-1-0079), and the Army Research Office MURI 
award W911NF0910406.
\end{acknowledgments}

\appendix
\section{Transfer cavity lock}
\label{laserlock}
We provide here further details on the transfer cavity lock, which is depicted in Fig.~\ref{LaserSetUp}:  The external cavity diode laser at 780 nm is locked to a Fabry-P\'{e}rot cavity using the Pound-Drever-Hall (PDH) technique~\cite{drever_laser_1983}.  The Fabry-P\'{e}rot cavity consists of plano and concave ($r=25$ cm) mirrors with high reflectivity (99.9\%) at both 780-nm and 842-nm, and the mirrors are attached to a fused silica spacer with a length of 10 cm.  In order to scan the resonance frequencies of the cavity, a piezoelectric transducer (PZT) ring is placed between one side of the spacer and the concave mirror.  With the measured cavity length, the corresponding free spectral range is 1.4 GHz, and the cavity has a linewidth of 1.4 MHz.  The cavity is isolated from acoustic vibration and placed inside a vacuum chamber.  The laser is locked to the cavity, which provides short time scale frequency stability, and then the cavity length is adjusted to bring the cavity and laser in resonance with a hyperfine transition of $^{87}$Rb.  Feeding back to the cavity PZT an error signal derived by Rb saturation absorption spectroscopy ensures that the cavity resonance frequencies drift by no more than a few hundred kHz.  

The cavity now provides a stable frequency reference for TiS1's 842 nm beam, which is picked-off from the main beam that is directed into the ring cavity frequency doubler.   To bridge the frequency gap between the nearest 842 nm mode and the doubled 842-nm wavelength necessary for generating the MOT light, the 842-nm beam passes through an optical fiber-based electro-optical modulator, which imprints strong sidebands at a frequency adjustable between 1.4 GHz--2.8 GHz.  A phase-locked loop generates this microwave frequency from a stabilized and tunable RF source.  One of the sidebands is adjusted in frequency to be resonant with the optical cavity, allowing the 842-nm to be locked to the cavity using the PDH technique.   Once locked, the 421-nm light from both TiS1 and TiS2 (which is offset beat note locked to TiS1) may be scanned to the correct frequency by adjusting the RF reference.


\begin{thebibliography}{99}%
\makeatletter
\providecommand \@ifxundefined [1]{%
 \ifx #1\undefined \expandafter \@firstoftwo
 \else \expandafter \@secondoftwo
\fi
}%
\providecommand \@ifnum [1]{%
 \ifnum #1\expandafter \@firstoftwo
 \else \expandafter \@secondoftwo
\fi
}%
\providecommand \enquote [1]{``#1''}%
\providecommand \bibnamefont  [1]{#1}%
\providecommand \bibfnamefont [1]{#1}%
\providecommand \citenamefont [1]{#1}%
\providecommand\href[0]{\@sanitize\@href}%
\providecommand\@href[1]{\endgroup\@@startlink{#1}\endgroup\@@href}%
\providecommand\@@href[1]{#1\@@endlink}%
\providecommand \@sanitize [0]{\begingroup\catcode`\&12\catcode`\#12\relax}%
\@ifxundefined \pdfoutput {\@firstoftwo}{%
 \@ifnum{\z@=\pdfoutput}{\@firstoftwo}{\@secondoftwo}%
}{%
 \providecommand\@@startlink[1]{\leavevmode}%
 \providecommand\@@endlink[0]{}%
}{%
 \providecommand\@@startlink[1]{%
  \leavevmode
  \pdfstartlink
   attr{/Border[0 0 1 ]/H/I/C[0 1 1]}%
   user{/Subtype/Link/A<</Type/Action/S/URI/URI(#1)>>}%
  \relax
 }%
 \providecommand\@@endlink[0]{\pdfendlink}%
}%
\providecommand \url  [0]{\begingroup\@sanitize \@url }%
\providecommand \@url [1]{\endgroup\@href {#1}{\urlprefix}}%
\providecommand \urlprefix [0]{URL }%
\providecommand \Eprint[0]{\href }%
\@ifxundefined \urlstyle {%
  \providecommand \doi [1]{doi:\discretionary{}{}{}#1}%
}{%
  \providecommand \doi [0]{doi:\discretionary{}{}{}\begingroup
  \urlstyle{rm}\Url }%
}%
\providecommand \doibase [0]{http://dx.doi.org/}%
\providecommand \Doi[1]{\href{\doibase#1}}%
\providecommand \bibAnnote [3]{%
  \BibitemShut{#1}%
  \begin{quotation}\noindent
    \textsc{Key:}\ #2\\\textsc{Annotation:}\ #3%
  \end{quotation}%
}%
\providecommand \bibAnnoteFile [2]{%
  \IfFileExists{#2}{\bibAnnote {#1} {#2} {\input{#2}}}{}%
}%
\providecommand \typeout [0]{\immediate \write \m@ne }%
\providecommand \selectlanguage [0]{\@gobble}%
\providecommand \bibinfo [0]{\@secondoftwo}%
\providecommand \bibfield [0]{\@secondoftwo}%
\providecommand \translation [1]{[#1]}%
\providecommand \BibitemOpen[0]{}%
\providecommand \bibitemStop [0]{}%
\providecommand \bibitemNoStop [0]{.\EOS\space}%
\providecommand \EOS [0]{\spacefactor3000\relax}%
\providecommand \BibitemShut [1]{\csname bibitem#1\endcsname}%
\bibitem{Fradkin2009}%
  \BibitemOpen
  \bibfield{author}{%
  \bibinfo {author} {\bibfnamefont{E.}~\bibnamefont{Fradkin}}, \bibinfo
  {author} {\bibfnamefont{S.~A.}\ \bibnamefont{Kivelson}}, \bibinfo {author}
  {\bibfnamefont{M.~J.}\ \bibnamefont{Lawler}}, \bibinfo {author}
  {\bibfnamefont{J.~P.}\ \bibnamefont{Eisenstein}},\ and\ \bibinfo {author}
  {\bibfnamefont{A.~P.}\ \bibnamefont{Mackenzie}},\ }%
  \bibfield{journal}{%
  \bibinfo {journal} {Annu. Rev. Condens. Matter Phys.}\ }%
  \textbf{\bibinfo {volume} {\textbf{1}}},\ \bibinfo {pages} {7.1} (\bibinfo
  {year} {2010})%
  \bibAnnoteFile{NoStop}{Fradkin2009}%
\bibitem{Fradkin2010}%
  \BibitemOpen
  \bibfield{author}{%
  \bibinfo {author} {\bibfnamefont{E.}~\bibnamefont{Fradkin}}\ and\ \bibinfo
  {author} {\bibfnamefont{S.~A.}\ \bibnamefont{Kivelson}},\ }%
  \bibfield{journal}{%
  \bibinfo {journal} {Science}\ }%
  \textbf{\bibinfo {volume} {327}},\ \bibinfo {pages} {155} (\bibinfo {year}
  {2010})%
  \bibAnnoteFile{NoStop}{Fradkin2010}%
\bibitem{Miyakawa:2008}%
  \BibitemOpen
  \bibfield{author}{%
  \bibinfo {author} {\bibfnamefont{T.}~\bibnamefont{Miyakawa}}, \bibinfo
  {author} {\bibfnamefont{T.}~\bibnamefont{Sogo}},\ and\ \bibinfo {author}
  {\bibfnamefont{H.}~\bibnamefont{Pu}},\ }%
  \bibfield{journal}{%
  \bibinfo {journal} {Phys. Rev. A}\ }%
  \textbf{\bibinfo {volume} {77}},\ \bibinfo {pages} {061603} (\bibinfo {year}
  {2008})%
  \bibAnnoteFile{NoStop}{Miyakawa:2008}%
\bibitem{Fregoso:2009}%
  \BibitemOpen
  \bibfield{author}{%
  \bibinfo {author} {\bibfnamefont{B.~M.}\ \bibnamefont{Fregoso}}, \bibinfo
  {author} {\bibfnamefont{K.}~\bibnamefont{Sun}}, \bibinfo {author}
  {\bibfnamefont{E.}~\bibnamefont{Fradkin}},\ and\ \bibinfo {author}
  {\bibfnamefont{B.~L.}\ \bibnamefont{Lev}},\ }%
  \bibfield{journal}{%
  \bibinfo {journal} {New J. Phys.}\ }%
  \textbf{\bibinfo {volume} {11}},\ \bibinfo {pages} {103003} (\bibinfo {year}
  {2009})%
  \bibAnnoteFile{NoStop}{Fregoso:2009}%
\bibitem{Quintanilla:2009}%
  \BibitemOpen
  \bibfield{author}{%
  \bibinfo {author} {\bibfnamefont{J.}~\bibnamefont{Quintanilla}}, \bibinfo
  {author} {\bibfnamefont{S.~T.}\ \bibnamefont{Carr}},\ and\ \bibinfo {author}
  {\bibfnamefont{J.~J.}\ \bibnamefont{Betouras}},\ }%
  \bibfield{journal}{%
  \bibinfo {journal} {Phys. Rev. A}\ }%
  \textbf{\bibinfo {volume} {79}},\ \bibinfo {pages} {031601} (\bibinfo {year}
  {2009})%
  \bibAnnoteFile{NoStop}{Quintanilla:2009}%
\bibitem{Fregoso2009b}%
  \BibitemOpen
  \bibfield{author}{%
  \bibinfo {author} {\bibfnamefont{B.~M.}\ \bibnamefont{Fregoso}}\ and\
  \bibinfo {author} {\bibfnamefont{E.}~\bibnamefont{Fradkin}},\ }%
  \bibfield{journal}{%
  \bibinfo {journal} {Phys. Rev. Lett.}\ }%
  \textbf{\bibinfo {volume} {103}},\ \bibinfo {pages} {205301} (\bibinfo {year}
  {2009})%
  \bibAnnoteFile{NoStop}{Fregoso2009b}%
\bibitem{fregoso:2010}%
  \BibitemOpen
  \bibfield{author}{%
  \bibinfo {author} {\bibfnamefont{B.~M.}\ \bibnamefont{Fregoso}}\ and\
  \bibinfo {author} {\bibfnamefont{E.}~\bibnamefont{Fradkin }}}%
   (\bibinfo {year} {2010}),\
  \Eprint{http://arxiv.org/abs/1001.4167}{arXiv:1001.4167}%
  \bibAnnoteFile{NoStop}{fregoso:2010}%
\bibitem{Doyle04EPJDreview}%
  \BibitemOpen
  \bibfield{author}{%
  \bibinfo {author} {\bibfnamefont{J.}~\bibnamefont{Doyle}}, \bibinfo {author}
  {\bibfnamefont{B.}~\bibnamefont{Friedrich}}, \bibinfo {author}
  {\bibfnamefont{R.}~\bibnamefont{Krems}},\ and\ \bibinfo {author}
  {\bibfnamefont{F.}~\bibnamefont{Masnou-Seeuws}},\ }%
  \bibfield{journal}{%
  \bibinfo {journal} {Eur. Phys. J. D}\ }%
  \textbf{\bibinfo {volume} {\textbf{31}}},\ \bibinfo {pages} {149} (\bibinfo
  {year} {(2004)})%
  \bibAnnoteFile{NoStop}{Doyle04EPJDreview}%
\bibitem{Ye:2009}%
  \BibitemOpen
  \bibfield{author}{%
  \bibinfo {author} {\bibfnamefont{S.}~\bibnamefont{Ospelkaus}} \emph{et~al.},\
  }%
  \bibfield{journal}{%
  \bibinfo {journal} {Faraday Discuss.}\ }%
  \textbf{\bibinfo {volume} {142}},\ \bibinfo {pages} {351} (\bibinfo {year}
  {2009})%
  \bibAnnoteFile{NoStop}{Ye:2009}%
\bibitem{Ospelkaus2010}%
  \BibitemOpen
  \bibfield{author}{%
  \bibinfo {author} {\bibfnamefont{S.}~\bibnamefont{Ospelkaus}}, \bibinfo
  {author} {\bibfnamefont{K.-K.}\ \bibnamefont{Ni}}, \bibinfo {author}
  {\bibfnamefont{D.}~\bibnamefont{Wang}}, \bibinfo {author}
  {\bibfnamefont{M.~H.~G.}\ \bibnamefont{de~Miranda}}, \bibinfo {author}
  {\bibfnamefont{B.}~\bibnamefont{Neyenhuis}}, \bibinfo {author}
  {\bibfnamefont{G.}~\bibnamefont{Quemener}}, \bibinfo {author}
  {\bibfnamefont{P.~S.}\ \bibnamefont{Julienne}}, \bibinfo {author}
  {\bibfnamefont{J.~L.}\ \bibnamefont{Bohn}}, \bibinfo {author}
  {\bibfnamefont{D.~S.}\ \bibnamefont{Jin}},\ and\ \bibinfo {author}
  {\bibfnamefont{J.}~\bibnamefont{Ye}},\ }%
  \bibfield{journal}{%
  \bibinfo {journal} {Science}\ }%
  \textbf{\bibinfo {volume} {327}},\ \bibinfo {pages} {853} (\bibinfo {year}
  {2010})%
  \bibAnnoteFile{NoStop}{Ospelkaus2010}%
\bibitem{Ni2010}%
  \BibitemOpen
  \bibfield{author}{%
  \bibinfo {author} {\bibfnamefont{K.~K.}\ \bibnamefont{Ni}}, \bibinfo {author}
  {\bibfnamefont{S.}~\bibnamefont{Ospelkaus}}, \bibinfo {author}
  {\bibfnamefont{D.}~\bibnamefont{Wang}}, \bibinfo {author}
  {\bibfnamefont{G.}~\bibnamefont{Quemener}}, \bibinfo {author}
  {\bibfnamefont{B.}~\bibnamefont{Neyenhuis}}, \bibinfo {author}
  {\bibfnamefont{M.~H.~G.}\ \bibnamefont{{de Miranda}}}, \bibinfo {author}
  {\bibfnamefont{J.~L.}\ \bibnamefont{Bohn}}, \bibinfo {author}
  {\bibfnamefont{J.}~\bibnamefont{Ye}},\ and\ \bibinfo {author}
  {\bibfnamefont{D.~S.}\ \bibnamefont{Jin}},\ }%
  \bibfield{journal}{%
  \bibinfo {journal} {Nature}\ }%
  \textbf{\bibinfo {volume} {464}},\ \bibinfo {pages} {3124} (\bibinfo {year}
  {2010})%
  \bibAnnoteFile{NoStop}{Ni2010}%
\bibitem{Pfau02}%
  \BibitemOpen
  \bibfield{author}{%
  \bibinfo {author} {\bibfnamefont{S.}~\bibnamefont{Giovanazzi}}, \bibinfo
  {author} {\bibfnamefont{A.}~\bibnamefont{{G\"{o}rlitz}}},\ and\ \bibinfo
  {author} {\bibfnamefont{T.}~\bibnamefont{Pfau}},\ }%
  \bibfield{journal}{%
  \bibinfo {journal} {Phys. Rev. Lett.}\ }%
  \textbf{\bibinfo {volume} {\textbf{89}}},\ \bibinfo {pages} {130401}
  (\bibinfo {year} {(2002)})%
  \bibAnnoteFile{NoStop}{Pfau02}%
\bibitem{Note1}%
  \BibitemOpen
  \bibinfo {note} {Terbium has only one isotope, a boson. Unfortunately, it
  possesses a 400 K electronic state that could be driven by incoherent
  blackbody radiation, thus limiting coherence and trap lifetimes.
  Additionally, thulium (4 $\mu _{B}$)~\cite {Thulium2010} and holmium (9 $\mu
  _{B}$) have only single bosonic isotopes.}%
  \bibAnnoteFile{Stop}{Note1}%
\bibitem{Lu2010}%
  \BibitemOpen
  \bibfield{author}{%
  \bibinfo {author} {\bibfnamefont{M.}~\bibnamefont{Lu}}, \bibinfo {author}
  {\bibfnamefont{S.-H.}\ \bibnamefont{Youn}},\ and\ \bibinfo {author}
  {\bibfnamefont{B.~L.}\ \bibnamefont{Lev}},\ }%
  \bibfield{journal}{%
  \bibinfo {journal} {Phys. Rev. Lett.}\ }%
  \textbf{\bibinfo {volume} {104}},\ \bibinfo {pages} {063001} (\bibinfo {year}
  {2010})%
  \bibAnnoteFile{NoStop}{Lu2010}%
\bibitem{Pfau05CrBEC}%
  \BibitemOpen
  \bibfield{author}{%
  \bibinfo {author} {\bibfnamefont{A.}~\bibnamefont{Griesmaier}}, \bibinfo
  {author} {\bibfnamefont{J.}~\bibnamefont{Werner}}, \bibinfo {author}
  {\bibfnamefont{S.}~\bibnamefont{Hensler}}, \bibinfo {author}
  {\bibfnamefont{J.}~\bibnamefont{Stuhler}},\ and\ \bibinfo {author}
  {\bibfnamefont{T.}~\bibnamefont{Pfau}},\ }%
  \bibfield{journal}{%
  \bibinfo {journal} {Phys. Rev. Lett.}\ }%
  \textbf{\bibinfo {volume} {\textbf{94}}},\ \bibinfo {pages} {160401}
  (\bibinfo {year} {(2005)})%
  \bibAnnoteFile{NoStop}{Pfau05CrBEC}%
\bibitem{PfauReview09}%
  \BibitemOpen
  \bibfield{author}{%
  \bibinfo {author} {\bibfnamefont{T.}~\bibnamefont{Lahaye}}, \bibinfo {author}
  {\bibfnamefont{C.}~\bibnamefont{Menotti}}, \bibinfo {author}
  {\bibfnamefont{L.}~\bibnamefont{Santos}}, \bibinfo {author}
  {\bibfnamefont{M.}~\bibnamefont{Lewenstein}},\ and\ \bibinfo {author}
  {\bibfnamefont{T.}~\bibnamefont{Pfau}},\ }%
  \bibfield{journal}{%
  \bibinfo {journal} {Rep. Prog. Phys.}\ }%
  \textbf{\bibinfo {volume} {72}},\ \bibinfo {pages} {126401} (\bibinfo {year}
  {2009})%
  \bibAnnoteFile{NoStop}{PfauReview09}%
\bibitem{Yi:2007}%
  \BibitemOpen
  \bibfield{author}{%
  \bibinfo {author} {\bibfnamefont{S.}~\bibnamefont{Yi}}, \bibinfo {author}
  {\bibfnamefont{T.}~\bibnamefont{Li}},\ and\ \bibinfo {author}
  {\bibfnamefont{C.}~\bibnamefont{Sun}},\ }%
  \bibfield{journal}{%
  \bibinfo {journal} {Phys. Rev. Lett.}\ }%
  \textbf{\bibinfo {volume} {98}},\ \bibinfo {pages} {260405} (\bibinfo {year}
  {2007})%
  \bibAnnoteFile{NoStop}{Yi:2007}%
\bibitem{Duan2010}%
  \BibitemOpen
  \bibfield{author}{%
  \bibinfo {author} {\bibfnamefont{Y.-H.}\ \bibnamefont{Chan}}, \bibinfo
  {author} {\bibfnamefont{Y.-J.}\ \bibnamefont{Han}},\ and\ \bibinfo {author}
  {\bibfnamefont{L.-M.}\ \bibnamefont{Duan }}}%
   (\bibinfo {year} {2010}),\
  \Eprint{http://arxiv.org/abs/1005.1270}{arXiv:1005.1270}%
  \bibAnnoteFile{NoStop}{Duan2010}%
\bibitem{Pfau07CrBECferrofluid}%
  \BibitemOpen
  \bibfield{author}{%
  \bibinfo {author} {\bibfnamefont{T.}~\bibnamefont{Lahaye}}, \bibinfo {author}
  {\bibfnamefont{T.}~\bibnamefont{Koch}}, \bibinfo {author}
  {\bibfnamefont{B.}~\bibnamefont{{Fr\"{o}hlich}}}, \bibinfo {author}
  {\bibfnamefont{M.}~\bibnamefont{Fattori}}, \bibinfo {author}
  {\bibfnamefont{J.}~\bibnamefont{Metz}}, \bibinfo {author}
  {\bibfnamefont{A.}~\bibnamefont{Griesmaier}}, \bibinfo {author}
  {\bibfnamefont{S.}~\bibnamefont{Giovanazzi}},\ and\ \bibinfo {author}
  {\bibfnamefont{T.}~\bibnamefont{Pfau}},\ }%
  \bibfield{journal}{%
  \bibinfo {journal} {Nature}\ }%
  \textbf{\bibinfo {volume} {\textbf{448}}},\ \bibinfo {pages} {672} (\bibinfo
  {year} {(2007)})%
  \bibAnnoteFile{NoStop}{Pfau07CrBECferrofluid}%
\bibitem{Note2}%
  \BibitemOpen
  \bibinfo {note} {Rapid three-body losses preclude greater reduction of
  $a_{s}$~\cite {Pfau07CrBECferrofluid}}%
  \bibAnnoteFile{NoStop}{Note2}%
\bibitem{Mcclelland:2006}%
  \BibitemOpen
  \bibfield{author}{%
  \bibinfo {author} {\bibfnamefont{J.~J.}\ \bibnamefont{McClelland}}\ and\
  \bibinfo {author} {\bibfnamefont{J.~L.}\ \bibnamefont{Hanssen}},\ }%
  \bibfield{journal}{%
  \bibinfo {journal} {Phys. Rev. Lett.}\ }%
  \textbf{\bibinfo {volume} {96}},\ \bibinfo {pages} {143005} (\bibinfo {year}
  {2006})%
  \bibAnnoteFile{NoStop}{Mcclelland:2006}%
\bibitem{FrenchCr06}%
  \BibitemOpen
  \bibfield{author}{%
  \bibinfo {author} {\bibfnamefont{R.}~\bibnamefont{Chicireanu}}, \bibinfo
  {author} {\bibfnamefont{A.}~\bibnamefont{Pouderous}}, \bibinfo {author}
  {\bibfnamefont{R.}~\bibnamefont{{Barb\'{e}}}}, \bibinfo {author}
  {\bibfnamefont{B.}~\bibnamefont{Laburthe-Tolra}}, \bibinfo {author}
  {\bibfnamefont{E.}~\bibnamefont{{Mar\'{e}chal}}}, \bibinfo {author}
  {\bibfnamefont{L.}~\bibnamefont{Vernac}}, \bibinfo {author}
  {\bibfnamefont{J.-C.}\ \bibnamefont{Keller}},\ and\ \bibinfo {author}
  {\bibfnamefont{O.}~\bibnamefont{Gorceix}},\ }%
  \bibfield{journal}{%
  \bibinfo {journal} {Phys. Rev. A}\ }%
  \textbf{\bibinfo {volume} {73}},\ \bibinfo {pages} {053406} (\bibinfo {year}
  {2006})%
  \bibAnnoteFile{NoStop}{FrenchCr06}%
\bibitem{Hancox:2004}%
  \BibitemOpen
  \bibfield{author}{%
  \bibinfo {author} {\bibfnamefont{C.}~\bibnamefont{Hancox}}, \bibinfo {author}
  {\bibfnamefont{S.}~\bibnamefont{Doret}}, \bibinfo {author}
  {\bibfnamefont{M.}~\bibnamefont{Hummon}}, \bibinfo {author}
  {\bibfnamefont{L.}~\bibnamefont{Luo}},\ and\ \bibinfo {author}
  {\bibfnamefont{J.}~\bibnamefont{Doyle}},\ }%
  \bibfield{journal}{%
  \bibinfo {journal} {Nature}\ }%
  \textbf{\bibinfo {volume} {431}},\ \bibinfo {pages} {281} (\bibinfo {year}
  {2004})%
  \bibAnnoteFile{NoStop}{Hancox:2004}%
\bibitem{Newman2010}%
  \BibitemOpen
  \bibfield{author}{%
  \bibinfo {author} {\bibfnamefont{B.}~\bibnamefont{Newman}}, \bibinfo {author}
  {\bibfnamefont{N.}~\bibnamefont{Brahms}}, \bibinfo {author}
  {\bibfnamefont{Y.-S.}\ \bibnamefont{Au}}, \bibinfo {author}
  {\bibfnamefont{C.}~\bibnamefont{Johnson}}, \bibinfo {author}
  {\bibfnamefont{C.}~\bibnamefont{Connolly}}, \bibinfo {author}
  {\bibfnamefont{J.~M.}\ \bibnamefont{Doyle}}, \bibinfo {author}
  {\bibfnamefont{D.}~\bibnamefont{Kleppner}},\ and\ \bibinfo {author}
  {\bibfnamefont{T.~J.}\ \bibnamefont{Greytak}}}%
   (\bibinfo {year} {2010}),\ \bibinfo {note} {to be published}%
  \bibAnnoteFile{NoStop}{Newman2010}%
\bibitem{Stamper-Kurn:2006}%
  \BibitemOpen
  \bibfield{author}{%
  \bibinfo {author} {\bibfnamefont{L.}~\bibnamefont{Sadler}}, \bibinfo {author}
  {\bibfnamefont{J.~M.}\ \bibnamefont{Higbie}}, \bibinfo {author}
  {\bibfnamefont{S.~R.}\ \bibnamefont{Leslie}}, \bibinfo {author}
  {\bibfnamefont{M.}~\bibnamefont{Vengalattore}},\ and\ \bibinfo {author}
  {\bibfnamefont{D.~M.}\ \bibnamefont{Stamper-Kurn}},\ }%
  \bibfield{journal}{%
  \bibinfo {journal} {Nature}\ }%
  \textbf{\bibinfo {volume} {443}},\ \bibinfo {pages} {312} (\bibinfo {year}
  {2006})%
  \bibAnnoteFile{NoStop}{Stamper-Kurn:2006}%
\bibitem{Wu:2006}%
  \BibitemOpen
  \bibfield{author}{%
  \bibinfo {author} {\bibfnamefont{C.}~\bibnamefont{Wu}},\ }%
  \bibfield{journal}{%
  \bibinfo {journal} {Mod. Phys. Lett. B}\ }%
  \textbf{\bibinfo {volume} {20}},\ \bibinfo {pages} {1707} (\bibinfo {year}
  {2006})%
  \bibAnnoteFile{NoStop}{Wu:2006}%
\bibitem{Baym09}%
  \BibitemOpen
  \bibfield{author}{%
  \bibinfo {author} {\bibfnamefont{G.}~\bibnamefont{Baym}}\ and\ \bibinfo
  {author} {\bibfnamefont{T.}~\bibnamefont{Hatsuda}},\ }%
  \bibinfo {note} {private communication (2009)}%
  \bibAnnoteFile{NoStop}{Baym09}%
\bibitem{CWu2010}%
  \BibitemOpen
  \bibfield{author}{%
  \bibinfo {author} {\bibfnamefont{Y.}~\bibnamefont{Li}}\ and\ \bibinfo
  {author} {\bibfnamefont{C.}~\bibnamefont{Wu }}}%
   (\bibinfo {year} {2010}),\
  \Eprint{http://arxiv.org/abs/1005.0889}{arXiv:1005.0889}%
  \bibAnnoteFile{NoStop}{CWu2010}%
\bibitem{Bohn10}%
  \BibitemOpen
  \bibfield{author}{%
  \bibinfo {author} {\bibfnamefont{S.}~\bibnamefont{Ronen}}\ and\ \bibinfo
  {author} {\bibfnamefont{J.~L.}\ \bibnamefont{Bohn}},\ }%
  \bibfield{journal}{%
  \bibinfo {journal} {Phys. Rev. A}\ }%
  \textbf{\bibinfo {volume} {81}},\ \bibinfo {pages} {033601} (\bibinfo {year}
  {2010})%
  \bibAnnoteFile{NoStop}{Bohn10}%
\bibitem{Bohn07}%
  \BibitemOpen
  \bibfield{author}{%
  \bibinfo {author} {\bibfnamefont{S.}~\bibnamefont{Ronen}}, \bibinfo {author}
  {\bibfnamefont{D.~C.~E.}\ \bibnamefont{Bortolotti}},\ and\ \bibinfo {author}
  {\bibfnamefont{J.~L.}\ \bibnamefont{Bohn}},\ }%
  \bibfield{journal}{%
  \bibinfo {journal} {Phys. Rev. Lett.}\ }%
  \textbf{\bibinfo {volume} {98}},\ \bibinfo {pages} {030406} (\bibinfo {year}
  {2007})%
  \bibAnnoteFile{NoStop}{Bohn07}%
\bibitem{Leefer:2009}%
  \BibitemOpen
  \bibfield{author}{%
  \bibinfo {author} {\bibfnamefont{N.~A.}\ \bibnamefont{Leefer}}, \bibinfo
  {author} {\bibfnamefont{A.}~\bibnamefont{Cing{\"o}z}}, \bibinfo {author}
  {\bibfnamefont{D.}~\bibnamefont{Budker}}, \bibinfo {author}
  {\bibfnamefont{S.~J.}\ \bibnamefont{Ferrell}}, \bibinfo {author}
  {\bibfnamefont{V.~V.}\ \bibnamefont{Yashchuk}}, \bibinfo {author}
  {\bibfnamefont{A.}~\bibnamefont{Lapierre}}, \bibinfo {author}
  {\bibfnamefont{A.~T.}\ \bibnamefont{Nguyen}}, \bibinfo {author}
  {\bibfnamefont{S.~K.}\ \bibnamefont{Lamoreaux}},\ and\ \bibinfo {author}
  {\bibfnamefont{J.~R.}\ \bibnamefont{Torgerson}}\ }%
  (\bibinfo {publisher} {World Scientific, Singapore},\ \bibinfo {year}
  {2009})\ pp.\ \bibinfo {pages} {34--43},\ \bibinfo {note} {{Proceedings of
  the 7th Symposium on Frequency Standards and Metrology}}%
  \bibAnnoteFile{NoStop}{Leefer:2009}%
\bibitem{Derevianko:2004}%
  \BibitemOpen
  \bibfield{author}{%
  \bibinfo {author} {\bibfnamefont{A.}~\bibnamefont{Derevianko}}\ and\ \bibinfo
  {author} {\bibfnamefont{C.}~\bibnamefont{Cannon}},\ }%
  \bibfield{journal}{%
  \bibinfo {journal} {Phys. Rev. A}\ }%
  \textbf{\bibinfo {volume} {70}},\ \bibinfo {pages} {062319} (\bibinfo {year}
  {2004})%
  \bibAnnoteFile{NoStop}{Derevianko:2004}%
\bibitem{Saffman:2008}%
  \BibitemOpen
  \bibfield{author}{%
  \bibinfo {author} {\bibfnamefont{M.}~\bibnamefont{Saffman}}\ and\ \bibinfo
  {author} {\bibfnamefont{K.}~\bibnamefont{M{\o}lmer}},\ }%
  \bibfield{journal}{%
  \bibinfo {journal} {Phys. Rev. A}\ }%
  \textbf{\bibinfo {volume} {78}},\ \bibinfo {pages} {012336} (\bibinfo {year}
  {2008})%
  \bibAnnoteFile{NoStop}{Saffman:2008}%
\bibitem{Connolly2010}%
  \BibitemOpen
  \bibfield{author}{%
  \bibinfo {author} {\bibfnamefont{C.~B.}\ \bibnamefont{Connolly}}, \bibinfo
  {author} {\bibfnamefont{Y.~S.}\ \bibnamefont{Au}}, \bibinfo {author}
  {\bibfnamefont{S.~C.}\ \bibnamefont{Doret}}, \bibinfo {author}
  {\bibfnamefont{W.}~\bibnamefont{Ketterle}},\ and\ \bibinfo {author}
  {\bibfnamefont{J.~M.}\ \bibnamefont{Doyle}},\ }%
  \bibfield{journal}{%
  \bibinfo {journal} {Phys. Rev. A}\ }%
  \textbf{\bibinfo {volume} {81}},\ \bibinfo {pages} {063001} (\bibinfo {year}
  {2010})%
  \bibAnnoteFile{NoStop}{Connolly2010}%
\bibitem{Zoller2010}%
  \BibitemOpen
  \bibfield{author}{%
  \bibinfo {author} {\bibfnamefont{M.}~\bibnamefont{Dalmonte}}, \bibinfo
  {author} {\bibfnamefont{G.}~\bibnamefont{Pupillo}},\ and\ \bibinfo {author}
  {\bibfnamefont{P.}~\bibnamefont{Zoller }}}%
   (\bibinfo {year} {2010}),\
  \Eprint{http://arxiv.org/abs/1004.5035}{arXiv:1004.5035}%
  \bibAnnoteFile{NoStop}{Zoller2010}%
\bibitem{Martin:1978}%
  \BibitemOpen
  \bibfield{author}{%
  \bibinfo {author} {\bibfnamefont{W.~C.}\ \bibnamefont{Martin}}, \bibinfo
  {author} {\bibfnamefont{R.}~\bibnamefont{Zalubas}},\ and\ \bibinfo {author}
  {\bibfnamefont{L.}~\bibnamefont{Hagan}},\ }%
  \emph{\bibinfo {title} {Atomic Energy Levels--The Rare Earth Elements}}\
  (\bibinfo {publisher} {NSRDS-NBS, \textbf{60}},\ \bibinfo {address}
  {Washington, D.C.},\ \bibinfo {year} {1978})%
  \bibAnnoteFile{NoStop}{Martin:1978}%
\bibitem{Leefer:2009b}%
  \BibitemOpen
  \bibfield{author}{%
  \bibinfo {author} {\bibfnamefont{N.}~\bibnamefont{Leefer}}, \bibinfo {author}
  {\bibfnamefont{A.}~\bibnamefont{Cing{\"o}z}},\ and\ \bibinfo {author}
  {\bibfnamefont{D.}~\bibnamefont{Budker}},\ }%
  \bibfield{journal}{%
  \bibinfo {journal} {Opt. Lett.}\ }%
  \textbf{\bibinfo {volume} {34}},\ \bibinfo {pages} {2548} (\bibinfo {year}
  {2008})%
  \bibAnnoteFile{NoStop}{Leefer:2009b}%
\bibitem{Hancox:2005}%
  \BibitemOpen
  \bibfield{author}{%
  \bibinfo {author} {\bibfnamefont{C.}~\bibnamefont{Hancox}},\ }%
  \bibinfo {note} {{Ph.D.} Thesis; Harvard University (2005)}%
  \bibAnnoteFile{NoStop}{Hancox:2005}%
\bibitem{Thulium2010}%
  \BibitemOpen
  \bibfield{author}{%
  \bibinfo {author} {\bibfnamefont{D.}~\bibnamefont{Sukachev}}, \bibinfo
  {author} {\bibfnamefont{A.}~\bibnamefont{Sokolov}}, \bibinfo {author}
  {\bibfnamefont{K.}~\bibnamefont{Chebakov}}, \bibinfo {author}
  {\bibfnamefont{A.}~\bibnamefont{Akimov}}, \bibinfo {author}
  {\bibfnamefont{S.}~\bibnamefont{Kanorsky}}, \bibinfo {author}
  {\bibfnamefont{N.}~\bibnamefont{Kolachevsky}},\ and\ \bibinfo {author}
  {\bibfnamefont{V.}~\bibnamefont{Sorokin }}}%
   (\bibinfo {year} {2010}),\
  \Eprint{http://arxiv.org/abs/1003.0877}{arXiv:1003.0877}%
  \bibAnnoteFile{NoStop}{Thulium2010}%
\bibitem{Lu2010b}%
  \BibitemOpen
  \bibfield{author}{%
  \bibinfo {author} {\bibfnamefont{M.}~\bibnamefont{Lu}} \emph{et~al.},\ }%
  \bibinfo {note} {in preparation}%
  \bibAnnoteFile{NoStop}{Lu2010b}%
\bibitem{Tekhnoscan}%
  \BibitemOpen
  \bibfield{author}{%
  \bibinfo {author} {\bibfnamefont{S.}~\bibnamefont{Kobtsev}}, \bibinfo
  {author} {\bibfnamefont{B.}~\bibnamefont{Lev}}, \bibinfo {author}
  {\bibfnamefont{J.}~\bibnamefont{Fortagh}},\ and\ \bibinfo {author}
  {\bibfnamefont{V.}~\bibnamefont{Baraulia}},\ }%
  \bibfield{journal}{%
  \bibinfo {journal} {Proc. of SPIE}\ }%
  \textbf{\bibinfo {volume} {7578}},\ \bibinfo {pages} {75782F} (\bibinfo
  {year} {2010})%
  \bibAnnoteFile{NoStop}{Tekhnoscan}%
\bibitem{schunemann_simple_1999}%
  \BibitemOpen
  \bibfield{author}{%
  \bibinfo {author} {\bibfnamefont{U.}~\bibnamefont{Schunemann}}, \bibinfo
  {author} {\bibfnamefont{H.}~\bibnamefont{Engler}}, \bibinfo {author}
  {\bibfnamefont{R.}~\bibnamefont{Grimm}}, \bibinfo {author}
  {\bibfnamefont{M.}~\bibnamefont{Weidemuller}},\ and\ \bibinfo {author}
  {\bibfnamefont{M.}~\bibnamefont{Zielonkowski}},\ }%
  \bibfield{journal}{%
  \bibinfo {journal} {Rev. Sci. Inst.}\ }%
  \textbf{\bibinfo {volume} {70}},\ \bibinfo {pages} {242} (\bibinfo {year}
  {1999})%
  \bibAnnoteFile{NoStop}{schunemann_simple_1999}%
\bibitem{Note3}%
  \BibitemOpen
  \bibinfo {note} {Custom made from SVT Associates, Inc.}%
  \bibAnnoteFile{Stop}{Note3}%
\bibitem{li_reduction_2002}%
  \BibitemOpen
  \bibfield{author}{%
  \bibinfo {author} {\bibfnamefont{D.}~\bibnamefont{Li}},\ }%
  \bibfield{journal}{%
  \bibinfo {journal} {J. Vac. Sci. Technol. A}\ }%
  \textbf{\bibinfo {volume} {20}},\ \bibinfo {pages} {33} (\bibinfo {year}
  {2002})%
  \bibAnnoteFile{NoStop}{li_reduction_2002}%
\bibitem{Leefer:2010}%
  \BibitemOpen
  \bibfield{author}{%
  \bibinfo {author} {\bibfnamefont{N.}~\bibnamefont{Leefer}}, \bibinfo {author}
  {\bibfnamefont{A.}~\bibnamefont{{Cing\"{o}z}}}, \bibinfo {author}
  {\bibfnamefont{B.}~\bibnamefont{Gerber-Siff}}, \bibinfo {author}
  {\bibfnamefont{A.}~\bibnamefont{Sharma}}, \bibinfo {author}
  {\bibfnamefont{J.~R.}\ \bibnamefont{Torgerson}},\ and\ \bibinfo {author}
  {\bibfnamefont{D.}~\bibnamefont{Budker}},\ }%
  \bibfield{journal}{%
  \bibinfo {journal} {Phys. Rev. A}\ }%
  \textbf{\bibinfo {volume} {81}},\ \bibinfo {pages} {043427} (\bibinfo {year}
  {2010})%
  \bibAnnoteFile{NoStop}{Leefer:2010}%
\bibitem{MetcalfBook99}%
  \BibitemOpen
  \bibfield{author}{%
  \bibinfo {author} {\bibfnamefont{H.~J.}\ \bibnamefont{Metcalf}}\ and\
  \bibinfo {author} {\bibnamefont{{P. van der Straten}}},\ }%
  \emph{\bibinfo {title} {Laser Cooling and Trapping}}\ (\bibinfo {publisher}
  {Springer-Verlag, New York},\ \bibinfo {year} {1999})%
  \bibAnnoteFile{NoStop}{MetcalfBook99}%
\bibitem{Note4}%
  \BibitemOpen
  \bibinfo {note} {These are design values, and have not been measured.}%
  \bibAnnoteFile{Stop}{Note4}%
\bibitem{Note5}%
  \BibitemOpen
  \bibinfo {note} {Unless stated, all the values of detuning are with respect
  to $^{164}$Dy atomic transition. Other isotopes require slightly modified
  detuning for optimal MOT performance.}%
  \bibAnnoteFile{Stop}{Note5}%
\bibitem{Note6}%
  \BibitemOpen
  \bibinfo {note} {All beam waists are reported as a beam $1/e^2$ radius.}%
  \bibAnnoteFile{Stop}{Note6}%
\bibitem{Note7}%
  \BibitemOpen
  \bibinfo {note} {The additional factor of 2.7 accounts for approximately
  isotropic polarization and equally distributed $m_{J}$'s in the MOT.}%
  \bibAnnoteFile{Stop}{Note7}%
\bibitem{Berglund:2007}%
  \BibitemOpen
  \bibfield{author}{%
  \bibinfo {author} {\bibfnamefont{A.~J.}\ \bibnamefont{Berglund}}, \bibinfo
  {author} {\bibfnamefont{S.~A.}\ \bibnamefont{Lee}},\ and\ \bibinfo {author}
  {\bibfnamefont{J.~J.}\ \bibnamefont{McClelland}},\ }%
  \bibfield{journal}{%
  \bibinfo {journal} {Phys. Rev. A}\ }%
  \textbf{\bibinfo {volume} {76}},\ \bibinfo {pages} {053418} (\bibinfo {year}
  {2007})%
  \bibAnnoteFile{NoStop}{Berglund:2007}%
\bibitem{Note8}%
  \BibitemOpen
  \bibinfo {note} {For Dy, $g_{g}=1.24$ and $g_{e}=1.22$.}%
  \bibAnnoteFile{Stop}{Note8}%
\bibitem{Andy2009}%
  \BibitemOpen
  \bibfield{author}{%
  \bibinfo {author} {\bibfnamefont{A.}~\bibnamefont{Berglund}}\ and\ \bibinfo
  {author} {\bibfnamefont{J.}~\bibnamefont{McClelland}},\ }%
  \bibinfo {note} {private communication (2009)}%
  \bibAnnoteFile{NoStop}{Andy2009}%
\bibitem{Flambaum2010}%
  \BibitemOpen
  \bibfield{author}{%
  \bibinfo {author} {\bibfnamefont{V.~A.}\ \bibnamefont{Dzuba}}\ and\ \bibinfo
  {author} {\bibfnamefont{V.~V.}\ \bibnamefont{Flambaum}},\ }%
  \bibfield{journal}{%
  \bibinfo {journal} {Phys. Rev. A}\ }%
  \textbf{\bibinfo {volume} {81}},\ \bibinfo {pages} {052515} (\bibinfo {year}
  {2010})%
  \bibAnnoteFile{NoStop}{Flambaum2010}%
\bibitem{Pfau01MT}%
  \BibitemOpen
  \bibfield{author}{%
  \bibinfo {author} {\bibfnamefont{J.}~\bibnamefont{Stuhler}}, \bibinfo
  {author} {\bibfnamefont{P.}~\bibnamefont{Schmidt}}, \bibinfo {author}
  {\bibfnamefont{S.}~\bibnamefont{Hensler}}, \bibinfo {author}
  {\bibfnamefont{J.}~\bibnamefont{Werner}}, \bibinfo {author}
  {\bibfnamefont{J.}~\bibnamefont{Mlynek}},\ and\ \bibinfo {author}
  {\bibfnamefont{T.}~\bibnamefont{Pfau}},\ }%
  \bibfield{journal}{%
  \bibinfo {journal} {Phys. Rev. A}\ }%
  \textbf{\bibinfo {volume} {\textbf{64}}},\ \bibinfo {pages} {031405(R)}
  (\bibinfo {year} {(2001)})%
  \bibAnnoteFile{NoStop}{Pfau01MT}%
\bibitem{Note9}%
  \BibitemOpen
  \bibinfo {note} {This is roughly the portion of the Zeeman slower in which
  the fast oven beam is in resonance with the slowing laser.}%
  \bibAnnoteFile{Stop}{Note9}%
\bibitem{Note10}%
  \BibitemOpen
  \bibinfo {note} {The transition strengths are nearly equal for the five
  stretched-state $\sigma _{+}$ transitions.}%
  \bibAnnoteFile{Stop}{Note10}%
  \bibitem{Neefer2010PC}%
  \BibitemOpen
  \bibfield{author}{%
  \bibinfo {author} {\bibfnamefont{N.}~\bibnamefont{Leefer}},\ }%
  \bibinfo {note} {private communication (2010)}%
  \bibAnnoteFile{NoStop}{Neefer2010PC}%
\bibitem{Bradley2000}%
  \BibitemOpen
  \bibfield{author}{%
  \bibinfo {author} {\bibfnamefont{C.~C.}\ \bibnamefont{Bradley}}, \bibinfo
  {author} {\bibfnamefont{J.~J.}\ \bibnamefont{{McClelland}}}, \bibinfo
  {author} {\bibfnamefont{W.~R.}\ \bibnamefont{Anderson}},\ and\ \bibinfo
  {author} {\bibfnamefont{R.~J.}\ \bibnamefont{Celotta}},\ }%
  \bibfield{journal}{%
  \bibinfo {journal} {Phys. Rev. A}\ }%
  \textbf{\bibinfo {volume} {61}},\ \bibinfo {pages} {053407} (\bibinfo {year}
  {2000})%
  \bibAnnoteFile{NoStop}{Bradley2000}%
\bibitem{Note11}%
  \BibitemOpen
  \bibinfo {note} {Including an $R_{\protect \text {lossMT}}$ loss term is
  equivalent to allowing $p$ to be non-unity, but the latter can be more
  physically motivated.}%
  \bibAnnoteFile{Stop}{Note11}%
\bibitem{Youn2010b}%
  \BibitemOpen
  \bibfield{author}{%
  \bibinfo {author} {\bibfnamefont{S.-H.}\ \bibnamefont{Youn}}, \bibinfo
  {author} {\bibfnamefont{M.}~\bibnamefont{Lu}},\ and\ \bibinfo {author}
  {\bibfnamefont{B.~L.}\ \bibnamefont{Lev}},\ }%
  \bibinfo {note} {in preparation}%
  \bibAnnoteFile{NoStop}{Youn2010b}%
\bibitem{Jhe:2004}%
  \BibitemOpen
  \bibfield{author}{%
  \bibinfo {author} {\bibfnamefont{K.}~\bibnamefont{Kim}}, \bibinfo {author}
  {\bibfnamefont{H.-R.}\ \bibnamefont{Noh}}, \bibinfo {author}
  {\bibfnamefont{H.-J.}\ \bibnamefont{Ha}},\ and\ \bibinfo {author}
  {\bibfnamefont{W.}~\bibnamefont{Jhe}},\ }%
  \bibfield{journal}{%
  \bibinfo {journal} {Phys. Rev. A}\ }%
  \textbf{\bibinfo {volume} {69}},\ \bibinfo {pages} {33406} (\bibinfo {year}
  {2004})%
  \bibAnnoteFile{NoStop}{Jhe:2004}%
\bibitem{Note12}%
  \BibitemOpen
  \bibinfo {note} {$\delta g_{\protect \text {Dy}}$ = 0.022 ($\delta
  g_{\protect \text {Dy}}/g_{\protect \text {Dy}}=1.7$\%), which is 5.5$\times
  $ larger than Er's on its MOT transition, but 7.7$\times $ less than Rb's.}%
  \bibAnnoteFile{Stop}{Note12}%
\bibitem{Berglund:2008}%
  \BibitemOpen
  \bibfield{author}{%
  \bibinfo {author} {\bibfnamefont{A.~J.}\ \bibnamefont{Berglund}}, \bibinfo
  {author} {\bibfnamefont{J.~L.}\ \bibnamefont{Hanssen}},\ and\ \bibinfo
  {author} {\bibfnamefont{J.~J.}\ \bibnamefont{McClelland}},\ }%
  \bibfield{journal}{%
  \bibinfo {journal} {Phys. Rev. Lett.}\ }%
  \textbf{\bibinfo {volume} {100}},\ \bibinfo {pages} {113002} (\bibinfo {year}
  {2008})%
  \bibAnnoteFile{NoStop}{Berglund:2008}%
\bibitem{ErTransitions05}%
  \BibitemOpen
  \bibfield{author}{%
  \bibinfo {author} {\bibfnamefont{H.~Y.}\ \bibnamefont{Ban}}, \bibinfo
  {author} {\bibfnamefont{M.}~\bibnamefont{Jacka}}, \bibinfo {author}
  {\bibfnamefont{J.~L.}\ \bibnamefont{Hanssen}}, \bibinfo {author}
  {\bibfnamefont{J.}~\bibnamefont{Reader}},\ and\ \bibinfo {author}
  {\bibfnamefont{J.~J.}\ \bibnamefont{McClelland}},\ }%
  \bibfield{journal}{%
  \bibinfo {journal} {Opt. Express}\ }%
  \textbf{\bibinfo {volume} {13}},\ \bibinfo {pages} {3186} (\bibinfo {year}
  {2005})%
  \bibAnnoteFile{NoStop}{ErTransitions05}%
\bibitem{Grimm09}%
  \BibitemOpen
  \bibfield{author}{%
  \bibinfo {author} {\bibfnamefont{R.}~\bibnamefont{Grimm}},\ }%
  \bibinfo {note} {private communication (2009)}%
  \bibAnnoteFile{NoStop}{Grimm09}%
\bibitem{drever_laser_1983}%
  \BibitemOpen
  \bibfield{author}{%
  \bibinfo {author} {\bibfnamefont{R.~W.~P.}\ \bibnamefont{Drever}}, \bibinfo
  {author} {\bibfnamefont{J.~L.}\ \bibnamefont{Hall}}, \bibinfo {author}
  {\bibfnamefont{F.~V.}\ \bibnamefont{Kowalski}}, \bibinfo {author}
  {\bibfnamefont{J.}~\bibnamefont{Hough}}, \bibinfo {author}
  {\bibfnamefont{G.~M.}\ \bibnamefont{Ford}}, \bibinfo {author}
  {\bibfnamefont{A.~J.}\ \bibnamefont{Munley}},\ and\ \bibinfo {author}
  {\bibfnamefont{H.}~\bibnamefont{Ward}},\ }%
  \bibfield{journal}{%
  \bibinfo {journal} {Appl. Phys. B}\ }%
  \textbf{\bibinfo {volume} {31}},\ \bibinfo {pages} {97} (\bibinfo {year}
  {1983})%
  \bibAnnoteFile{NoStop}{drever_laser_1983}%
\end{thebibliography}
%

\end{document}